\documentclass[amsmath,amssymb,amsbsy,prb,twocolumn,preprintnumbers,superscriptaddress,showpacs]{revtex4-1}
\usepackage[dvipdfmx]{graphicx,color}
\usepackage{dcolumn}
\usepackage{bm}
\usepackage{braket}
\usepackage{mathtools}
\usepackage{ulem}
\usepackage[breaklinks,colorlinks=true,linkcolor=blue,urlcolor=blue,citecolor=blue]{hyperref}

\begin{document}
\title{Trimer classical spin liquid from interacting fractional charges}
\author{Kunio Tokushuku}
\affiliation{Department of Physics, University of Tokyo,
Hongo, Bunkyo-ku, Tokyo 113-0033, Japan }
 \email{tokushuku@hosi.phys.s.u-tokyo.ac.jp}

\author{Tomonari Mizoguchi}
\affiliation{Department of Physics, University of Tsukuba, Tsukuba, Ibaraki 305-8571, Japan}

\author{Masafumi Udagawa}
\affiliation{Department of Physics, Gakushuin University, Mejiro, Toshima-ku, Tokyo 171-8588, Japan}
\affiliation{Max-Planck-Institut f\"{u}r Physik komplexer Systeme, 01187 Dresden, Germany}

\begin{abstract}
We study a problem of interacting fractional charges with $J_1$-$J_2$-$J_3$ Ising model on a checkerboard lattice under magnetic field.
As a result of the interplay between repulsive interactions and particle density tuning by magnetic field, the fractional charges form a classical spin liquid (CSL) phase.
The CSL phase is composed of degenerate spin configurations, which can be mapped to the trimer covering of dual square lattice.
The CSL state shows macroscopic ground-state entropy, implying the emergence of novel quantum spin liquid phase when quantum fluctuations are turned on.
In addition to the CSL phase, the system exhibits multiple magnetization plateaus, reflecting the fertile screening processes of dimer-monomer mixtures.
\end{abstract}

\maketitle

\section{Introduction}
Fractionalization is one of the central topics in condensed matter physics. In many-body interacting systems, low-energy excitations are usually described as quasiparticles, whose nonuniversal parameters, such as mass, have renormalized values, while their quantum numbers are preserved to be the same as those of the original particles. Fractionalization changes this canonical description considerably, by allowing the particles to split into subunits with smaller quantum numbers.

Fractional excitations are hosted in a number of systems, such as one-dimensional quantum liquid\cite{Bethe1931,mourigal2013fractional,PhysRevB.41.2326,kitaev2001unpaired}, fractional quantum Hall systems\cite{laughlin1983anomalous,tsui1982two}, and quantum spin liquids (QSLs)~\cite{kitaev2006anyons,knolle2014dynamics,doi:10.1146/annurev-conmatphys-020911-125058,balents2010balents}.
Among them, in the candidate compounds of QSLs, the role of fractional excitations has been highlighted in their thermodynamic and transport behaviors\cite{tokiwa2016tokiwa,pan2016measure,PhysRevX.1.021002,kimura2013quantum,Sibille2018,PhysRevLett.122.117201,wan2016spinon,kourtis2016free,PhysRevLett.120.167202,PhysRevB.96.195127,banerjee2016proximate,banerjee2017neutron,2017NatPh..13.1079D,kasahara2018majorana,knolle2014dynamics}.
In particular, the relevant role of fractional excitations is established in classical spin liquids (CSLs).
CSL corresponds to a high-temperature precursor of QSL, composed of a macroscopic number of degenerate classical states.
As lowering temperatures, coherency develop among the classical states, turning CSL into QSL.
One can find several essential properties of QSL already show up in CSL. 
Among other things, fractional excitations can be defined in CSL, usually as static objects.

While the appearance of fractionalization itself requires interaction between original particles, the assembly of fractional particles also hosts nontrivial many-body problems.
The problem of interacting fractional particles is usually quite difficult to treat theoretically, however, at the level of CSL, rigorous theoretical treatments are sometimes possible.
In a class of CSLs defined on frustrated magnets, we can clearly divide the role of interactions: the nearest-neighbor interaction leads to the formation of CSL with emergent fractional excitations,
while the farther-neighbor interactions give rise to the interaction between the fractional particles.
A typical example can be found in spin ice, a material realization of three-dimensional Coulomb phase, where a fractionalized particle, called monopole, dominates the low-energy properties of the system\cite{castelnovo2008magnetic}.

Indeed, in dipolar spin ice, monopoles exhibit interesting many-body effects, due to their Coulombic attractive interactions proportional to $-1/r$\cite{castelnovo2008magnetic}.
This Coulomb attraction 
gives rise to a liquid-gas-type phase transition elusive in a magnetic system\cite{PhysRevLett.90.207205}.
This phase transition is controlled by magnetic field as a tuning parameter, which is translated into a chemical potential of monopoles.
This analogy naturally leads to the plasma description of long-distance physics, where the screening of charge plays an important role in the thermodynamic behavior of the system.

It is also interesting to turn a look at fractional particles under short-range interactions.
In this case, the system is more susceptible to the hardcore constraint arising from the local charge conservation. 
In a variant of spin-ice-type frustrated magnet, described by $J_1$-$J_2$-$J_3$ Ising model, again the nearest-neighbor interaction ($J_1$) leads to the formation of CSL,
and the second- to third-neighbor interactions ($J_2, J_3$) give rise to the short-range interactions between them\cite{PhysRevLett.119.077207}.
In this system, the fractional excitations have magnetic charges as well-defined quantum numbers, and they satisfy a conservation law. 
In conventional electromagnetism, same-sign charges repel each other.
Meanwhile in this emergent Coulomb phase, same-sign charges sometimes attract each other.
However, the conservation law strictly forbids the same-sign charges to pair-annihilate.
This keen competition between the conservation law and the interactions leads to the formation of novel classical spin liquids\cite{PhysRevLett.119.077207,PhysRevB.94.104416,rau2016spin}, 
excitations \cite{mizoguchi2018magnetic,PhysRevB.98.140402}, and dynamics \cite{PhysRevB.94.104416}.

The stream of results motivates us to look into the magnetization process of this CSL.
Rich behaviors of the magnetization process have been explored so far in magnetic systems, especially with geometrical frustration\cite{PhysRevB.67.104431,PhysRevB.96.180401,PhysRevLett.102.137201,PhysRevB.94.075136,PhysRevB.62.15067,PhysRevLett.112.127203,PhysRevB.83.100405,PhysRevLett.108.057205,PhysRevLett.110.267201,doi:10.1143/JPSJ.69.1016,PhysRevLett.111.137204,PhysRevB.87.214424,doi:10.7566/JPSJ.83.113703}.
In general, the magnetic field affects the monopoles as a staggered potential and facilitates the pair-creation of monopoles.
The interaction between the induced monopoles may lead to a formation of novel CSL with macroscopic degeneracy.
In fact, a topologically ordered state is proposed at a low-field plateau of the kagome Heisenberg magnet\cite{nishimoto2013controlling}.

In this paper, we investigate a magnetic phase diagram and magnetization processes of the $J_1$-$J_2$-$J_3$ Ising model on a checkerboard lattice,
which is a two-dimensional analog of the pyrochlore lattice.
We obtain a ground state phase diagram analytically up to small positive farther-neighbor interactions,  
and find that there is a series of exotic phases which are characterized by magnetic plateaus. 
Remarkably, at the $1/3$-magnetization plateau, we find a CSL phase, where the whole system is tiled with magnetic trimers.
This state is a new-type of CSL with novel value of residual entropy.
It implies that a new class of QSL's emerges upon adding perturbations to this CSL state.

The rest of this paper is organized as follows. 
In Sec.~\ref{sec:model}, we present our model and methods, with a special focus on Gauss' law, which is central to the rigorous argument we base the existence of the CSL on.
In Sec.~\ref{sec:Phase diagram}, we show the overall ground-state phase diagram and present the results at simple limiting cases.
In Sec.~\ref{sec:tCSL}, we show the existence of the trimer classical spin liquid from both the intuitive and rigorous arguments, and detail its properties.
In Sec.~\ref{sec:magnetization}, we consider a full magnetization process and show a variety of commensurate phases, which are characterized by magnetic plateaus at
$M=0, 1/5, 1/3, 1/2, 2/3, 3/4$ and $1$. 
Finally, in Sec.~\ref{sec:summary}, we summarize our results and provide some future perspectives.

\section{Model and Method  \label{sec:model}}
\subsection{Model}
\begin{figure}[]
\begin{center}
\includegraphics[width=\hsize]{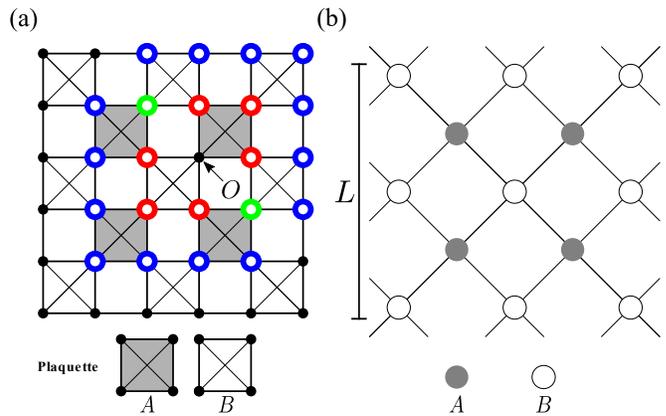}
\end{center}
\caption{(a) A schematic picture of checkerboard lattice. The definition of $J_1$, $J_2$ and $J_3$ couplings are shown here:
the site $O$ connects with red, green and blue sites with interactions, $J_1$, $J_2$, and $J_3$, respectively.
The plaquettes are divided into two sublattices. The shaded (white) plaquettes belong to the sublattice A (B).
(b) A dual square lattice obtained by replacing a plaquette with a site, and by connecting neighboring plaquettes with a bond.
Shaded (blank) sites correspond to the plaquettes on the sublattice A (B). Magnetic charge is defined on each site of this dual lattice, and neighboring charges interact through the coupling $J$, as defined in Eq.~(\ref{eq:defJ}). 
}
\label{fig:interaction}
\end{figure}
We consider the $J_1$-$J_2$-$J_3$ Ising model on a checkerboard lattice (Fig.~\ref{fig:interaction}).
The checkerboard lattice is a corner-sharing network of square units with diagonal bonds, which we simply call ``plaquettes".
We consider the network composed of $N_p$ plaquettes, i.e., $2N_p$ spins, in a periodic boundary condition in both directions.
The Hamiltonian of our model is defined as
\begin{align}
\mathcal{H}=& J_1 \sum_{\langle i,j \rangle_{\rm{n.n.}}} \sigma^{z}_i\sigma^{z}_j
+J_2 \sum_{ \langle i,j \rangle_{\rm{2nd}}} \sigma^{z}_i\sigma^{z}_j \nonumber \\
&+J_3\sum_{\langle i,j \rangle_{\rm{3rd}}} \sigma^{z}_i\sigma^{z}_j - h\sum_{i}\sigma^{z}_i, \label{eq:hamiltonian}
\end{align} 
where $\sigma^{z}_i=\pm1$ are Ising spins and $h$ represents an external magnetic field. Throughout this paper, we consider the case of $h\geq0$ without loss of generality.
We assume the nearest-neighbor (n.n.) coupling as antiferromagnetic and set its value as a unit energy ($J_1$=1). 
The n.n. coupling accounts for the interaction within the plaquettes.
In addition to $J_1$, we introduce $J_2$ as the diagonal interaction, and $J_3$ as the other next-nearest-neighbor interactions, as shown in Fig.~\ref{fig:interaction}.
Throughout the paper, we fix the ratio of $J_2$ and $J_3$ as
\begin{align}
J_2= 2J,\ \ J_3=J. \label{eq:defJ}
\end{align}
This point is of special importance, since the model can be mapped to the Hamiltonian of interacting magnetic charges with a staggered potential.
To describe the mapping, we first introduce a magnetic charge:
\begin{align}
Q_p=\eta_p S_p,\label{eq:defcharge}
\end{align}
where $p$ is a label of plaquettes and we define the total spin of the plaquette $p$, as
\begin{align}
S_p=\sum_{i\in p}\sigma^{z}_i,\label{eq:deftotalspin}
\end{align}
and the sign factor of the plaquette as
\begin{align}
\eta_p=
\{ 
\begin{array}{ll}
1 & (\mathrm{for}\; A\; \mathrm{sub.}) \\
-1 & (\mathrm{for}\; B\; \mathrm{sub.})
\end{array}.  
\end{align}
Here, we took the dual picture and regarded the checkerboard lattice as a square lattice of plaquettes,
and made the bipartite decomposition of the plaquettes into sublattice $A$ and $B$ as shown in Fig.~\ref{fig:interaction}. 
Note that there are five possible values of $Q_p$: $Q_p=0, \pm2, \pm4$ (Fig.~\ref{fig:TopoCharge}). Among them, we especially call the plaquette with $Q_p=0$, a vacuum plaquette, 
and that with $|Q_p|=4$ as a double charge. Since the neighboring plaquettes share one spin, the charges are not independent of each other.
Indeed, the magnetic charge has a conserved nature, and its distribution is a subject to the lattice analog of Gauss' law, as we will show later.

In terms of $Q_p$, we can rewrite the Hamiltonian~(\ref{eq:hamiltonian}) as
\begin{align}
\mathcal{H}=\left( \frac{1}{2} -J \right) \sum_{p}Q_p^2-J\sum_{\langle p,q \rangle}Q_pQ_q 
- \frac{h}{2}\sum_p\eta_pQ_p + C_1, \label{eq:DumbbelHamiltonianMag}
\end{align}
where 
\begin{align}
C_1=2(J-1)N_p.
\end{align}
We show the detail of this transformation in Appendix A.
In Eq.~(\ref{eq:DumbbelHamiltonianMag}), the first term is a ``self-energy" of the charges which is proportional to $Q_p^2$.
The second term is the interaction between magnetic charges on nearest-neighbor plaquettes.
The coupling constant of this interaction is $J$.
Accordingly, same-sign charges repel (attract) each other for $J<0$ ($J>0$).
The third term is a staggered potential for charges, which arises from the magnetic field.

\subsection{Gauss' law}
\begin{figure}[]
\begin{center}
\includegraphics[width=0.7\hsize]{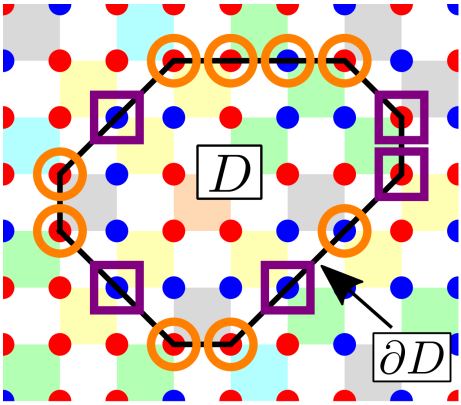}
\end{center}
\caption{
Schematic picture of the notation used to describe Gauss' law Eq.~(\ref{eq:gausslaw}). 
The black line represents the boundary, $\partial D$, and the region inside corresponds to the domain, $D$.
On the boundary, the sites with $\eta_{p_{D(i)}}=+1 (-1)$ are marked with circles (squares).
We have seven up and two down spins marked with circles, while two up and three down spins marked with squares.
Accordingly, we have $\sum_{i \in \partial D}\eta_{p_{D(i)}}\sigma^z_i=7\times(+1)+(-2)\times(+1)+2\times(-1)+(-3)\times(-1)=6$.
The domain $D$ involves one plaquette with $Q_p=+4$, four plaquettes with $Q_p=+2$, three plaquettes with $Q_p=0$, and three plaquettes with $Q_p=-2$.
Accordingly, we obtain $\sum_{p\in D}Q_p=1\times(+4)+4\times(+2)+3\times(0)+3\times(-2)=6$, which is equal to $\sum_{i \in \partial D}\eta_{p_{D(i)}}\sigma^z_i$.
}
\label{fig:gausslaw}
\end{figure}

The magnetic charge, $Q_p$, satisfies the conservation law. 
To see this, suppose that $D$ is an arbitrary set of plaquettes and $\partial D$ denotes the boundary sites of $D$ (see Fig.~\ref{fig:gausslaw} ).  
Then, the total magnetic charges inside $D$ and the spins on $\partial D$
satisfy the following relation: 
\begin{align}
\sum_{p\in D}Q_p=\sum_{i \in \partial D}\eta_{p_{D(i)}}\sigma^z_i. \label{eq:gausslaw}
\end{align}
Here, the boundary site
belongs to the two plaquettes, one inside, and one outside $D$, and $p_{D(i)}$ stands for the former.
Equation (\ref{eq:gausslaw}) means that the total charge in a certain region equals the sum of ``flux", $\eta_{p_{D(i)}}\sigma^z_i$, on its boundary.
This is nothing but a lattice analog of Gauss' law.

This Gauss' law constrains the structure of charge configuration in the system. In particular, as developed in Ref.~\onlinecite{PhysRevLett.119.077207}, this law leads to the triangle inequality:
\begin{align}
|\sum_{p\in D}Q_p|\leq\sum_{i \in \partial D}|\eta_{p_{D(i)}}\sigma^z_i|=N_{\partial D}. \label{eq:gaussinequality}
\end{align}
Namely, the amount of total charge an arbitrary region $D$ can store is bounded by $N_{\partial D}$, the number of its boundary sites.

\section{Phase diagram  \label{sec:Phase diagram}}
\begin{figure}[]
\begin{center}
\includegraphics[width=\hsize]{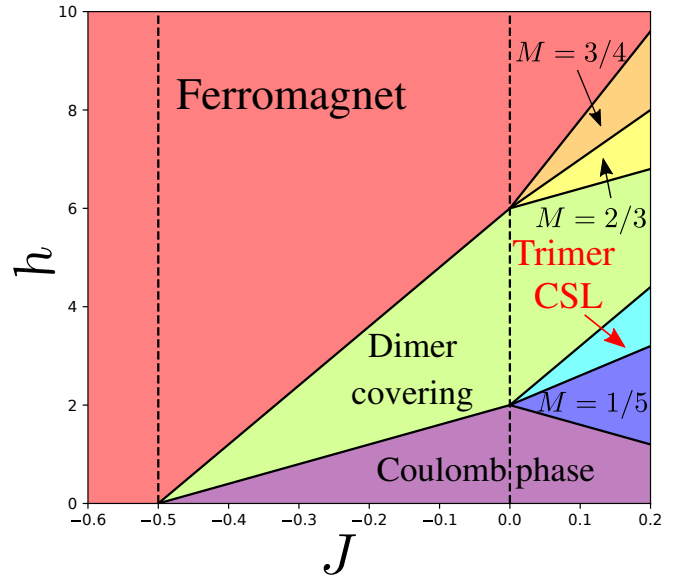}
\end{center}
\caption{The ground state phase diagram of the Hamiltonian, Eq.~(\ref{eq:DumbbelHamiltonianMag}), up to small positive $J$, obtained by the analytical arguments in Secs.~\ref{sec:tCSL} and \ref{sec:magnetization}.
}
\label{fig:PhaseDiagram}
\end{figure}

Before going into the details of analyses, we first show the overall phase diagram of the present model in Fig.~\ref{fig:PhaseDiagram}.
Each phase corresponds to a magnetization plateau, which we characterize by the magnetization per spin, $M\equiv\frac{1}{2N_p}\sum_{i}\sigma_i^z$.
Here, we focus on the region up to small positive $J$. 

In this section, we consider the simple limiting cases, and first discuss the stability of the Coulomb phase at $h=0$, then consider the magnetization process for $J\leq0$.
We also give a brief introduction to the ground states for $J>1/4$ and $J<-1/2$.

\begin{figure}[]
\begin{center}
\includegraphics[width=\hsize]{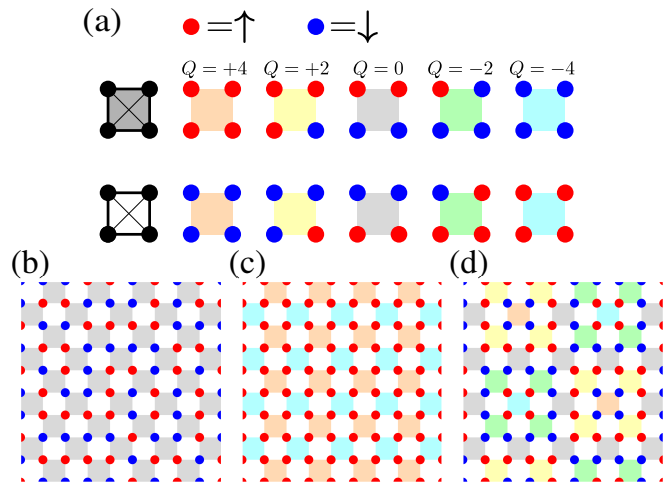}
\end{center}
\caption{(a) The relation between the value of magnetic charge and spin configurations on each sublattice. The red and blue circles mean up and down spins, respectively. Colors of plaquettes represent values of magnetic charges.
(b)-(d) The schematic picture of ground state spin configurations in the absence of magnetic field: (b) $-1/2<J<1/4$, (c) $J<-1/2$ and (d) $J>1/4$.}
\label{fig:TopoCharge}
\end{figure}

\subsection{Zero-field states \label{sec:zerofield}}
Let us start with the ground states in the absence of magnetic field.
Without magnetic field, Hamiltonian~(\ref{eq:DumbbelHamiltonianMag}) is simplified to
\begin{align}
\mathcal{H}=\left( \frac{1}{2}-J \right) \sum_{p}Q_p^2-J\sum_{\langle p,q \rangle}Q_pQ_q \label{eq:DumbbelHamiltonian}.
\end{align}

\subsubsection{$J=0$: square ice}
For $J=0$, where there is no interaction between charges, the model is reduced to the square-ice model only with n.n. exchange interactions
whose ground state is given by the charge vacuum ($Q_p = 0$ for every $p$), or the two-dimensional Coulomb phase [Fig.~\ref{fig:TopoCharge}(b)], which is a typical CSL state with macroscopic degeneracy.
Lieb's rigorous argument gives the exact value of residual entropy as 
$\mathcal{S}_0^{\rm SI}=\frac{1}{2}\ln W=0.215566427$ per spin, with $W=\frac{4}{3}^{\frac{3}{2}}$~\cite{PhysRevLett.18.692}.

\subsubsection{$J<0$: staggered charge ordering}
\label{sec:negativeJ_nofield}
For $J<0$, where an attractive interaction acts between opposite charges, 
the first term in Eq.~(\ref{eq:DumbbelHamiltonian}) favors a vacuum, whereas the second term favors a staggered charge ordering.
To see this competition clearly, we transform Eq.~(\ref{eq:DumbbelHamiltonian}) into
\begin{align}
\mathcal{H}=\left(\frac{1}{2}+J\right)\sum_{p}Q_p^2+\frac{|J|}{2}\sum_{\langle p,q \rangle}(Q_p+Q_q)^2. \label{eq:DumbbelHamiltonian2}
\end{align} 
With this form, one can minimize the first and the second terms simultaneously, by setting $Q_p = -Q_q$ for $p\in$ A sub. and $q\in$ B sub.,
and $|Q_p|= 0\ (4)$ for $J > -1/2$ ($J < -1/2$). This solution means the Coulomb phase extends to $J = -1/2$, while the staggered magnetic charge ordering with $|Q_p| = 4$ takes over for $J < -1/2$.
In the spin language, the staggered charge ordering with $|Q_p| = 4$ corresponds to the fully-polarized ferromagnetic state [Fig.~\ref{fig:TopoCharge}(c)].

\subsubsection{$J > 0$: long-period phase}
In contrast to the case of $J<0$, same-sign charges attract with each other for $J>0$. 
As a result, the Coulomb phase survives only up to $J=1/4$, and then the system turns to a complicated long-period spin ordering for $J>1/4$,
as in the case of pyrochlore model\cite{PhysRevB.94.104416}.

To obtain an insight into the ground state for positive $J$, let us rewrite the Hamiltonian Eq.~(\ref{eq:DumbbelHamiltonian}) as 
\begin{align}
\mathcal{H}=\left(\frac{1}{2}-3J\right)\sum_{p}Q_p^2+\frac{J}{2}\sum_{\langle p,q \rangle}(Q_p-Q_q)^2. 
\end{align} 
This expression shows the Coulomb phase is stable at least up to $J=1/6$ from a similar argument of $J<0$.
It also implies proliferation of double charges at larger $J$.
At first sight, it is preferable to cover the system with same signs of double charges for large $J$, however, according to the Gauss' law, the same-sign-charge cluster accommodates
at most one double charge \cite{PhysRevLett.119.077207}. As a result, the ordered phase contains mixed values of charges as shown in Fig.~\ref{fig:TopoCharge}(d). 
The precise value of the critical point of $J=1/4$ can be obtained by the softening of the lowest excitations, in the same argument as in pyrochlore model \cite{PhysRevB.94.104416}.
The formal proof based on the Gauss' law is rather involved. We summarize it in Appendix B.

\subsection{Magnetization process at $J\leq 0$}
\label{sec:Mprocess_negativeJ}
Next, let us look at the magnetization process for $J\leq 0$.
In this case, simple and rigorous arguments are available for the magnetization process, including the simplest non-interacting limit, $J=0$.
In this region, it is convenient to rewrite the Hamiltonian~(\ref{eq:DumbbelHamiltonianMag}) as
\begin{align}
\mathcal{H}=\left(\frac{1}{2} + J\right) \sum_p(Q_p-\eta_p S_h)^2-\frac{J}{2}\sum_{\langle p,q \rangle}\left( Q_p+Q_q \right)^2 + C_2,\label{eq:hhamiltonian2}
\end{align}
with $C_2=\frac{N_p}{8}\bigl[16J-16-\frac{h^2}{1+2J}\bigr].$
Here, we define
\begin{align}
S_h\equiv\frac{h/2}{1+2J},
\end{align} 
Then, in analogy with the argument in Sec.~\ref{sec:negativeJ_nofield}, 
the ground state can be obtained by minimizing the two terms in Eq.~(\ref{eq:hhamiltonian2}), simultaneously.
It is possible by minimizing $|Q_p-\eta_pS_h|$ while keeping the staggered charge alignment $Q_p+Q_q=0$ for any neighboring plaquettes, $p$ and $q$.

\subsubsection{Square ice: $0<h<2 + 4J$}
For small $h$, we find that the zero-field square ice, or the Coulomb phase, extends in the region: $S_h<1$ ($h<2 + 4J$).

\subsubsection{Dimer phase: $2  +4J <h<6  +12J$}
\begin{figure}[]
\begin{center}
\includegraphics[width=\hsize]{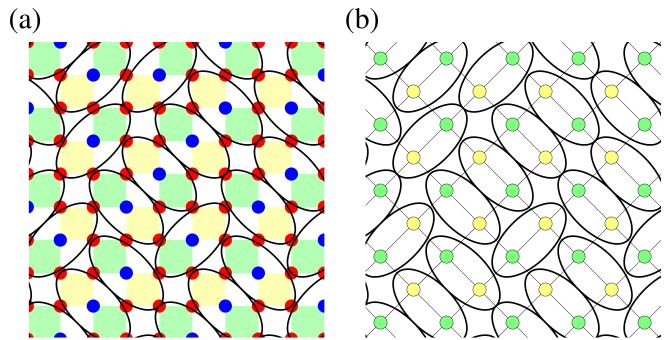}
\end{center}
\caption{(a) A schematic figure of spin configuration on the half-magnetization plateau and its dimer mapping. The yellow and green plaquettes support the magnetic charge, $Q_p=+2 $ and $-2$, as introduced in Fig.~\ref{fig:TopoCharge}. (b) The dimer configuration on a dual square lattice,
corresponding to the spin configuration in (a).
}
\label{fig:Kagomeice_mapping}
\end{figure}

For $S_h\geq 1$, the system goes out of the Coulomb phase, and the staggered charge ordering appears.
This region is divided into two cases.
First, for $1<S_h<3$, (i.e., $2  +4J <h<6  +12J$), $Q_p$ is given by 
\begin{align}
Q_p=
\{
\begin{array}{ll}
+2 & \mathrm{for}\; A\ \mathrm{sub.} \\
-2 & \mathrm{for}\; B\ \mathrm{sub.} 
\end{array}.
\end{align}
This phase corresponds to the half-magnetization plateau with $M=1/2$. While this phase has the staggered charge ordering,
it still keeps macroscopic degeneracy in spin degrees of freedom, forming a CSL state.
Its ground-state manifold can be mapped to the dimer covering problem, in a similar way to kagome ice\cite{matsuhira2002new,doi:10.1143/JPSJ.71.2365,isakov2004magnetization}.
To see this, let us consider a dual square lattice again, and place a dimer on each down spin.
Then, each spin configuration on this half-magnetization plateau can be mapped to a dimer configuration on the dual square lattice (Fig.~\ref{fig:Kagomeice_mapping}).
The residual entropy can be exactly obtained to be $\mathcal{S}_0^{\rm hmp}=0.14578045$ per spin by counting the dimer configuration on a square lattice \cite{PhysRev.124.1664}.
\subsubsection{The fully-polarized state: $6+12J < h$}
Finally, for $3 < S_h$, ($6+12J < h$), $Q_p$ is given by
\begin{align}
Q_p= \{ \begin{array}{ll}
+4 & \mathrm{for}\; A\ \mathrm{sub.} \\
-4 & \mathrm{for}\; B\ \mathrm{sub.} 
\end{array},
\end{align}
which corresponds to a fully-polarized ferromagnet. 

\section{Trimer CSL}
\label{sec:tCSL}
At $J=0$, the system makes a direct transition from the zero-field square ice phase to the dimer phase at $h=2$.
At $J>0$, between these two CSL's, there appears another type of CSL, which we call ``trimer CSL" after its structure.
In this section, we will address how the dimer phase makes instability to this trimer CSL, by the two kinds of strategies: one is based on the instability analysis due to the creation of vacuum plaquette, which is more intuitive and gives a clearer picture of this trimer CSL, and the other is a rigorous argument based on Gauss' law.

\subsection{Instability of the dimer phase}
\subsubsection{Nucleation of vacuum plaquette}
We start with the instability analysis of the dimer phase, as the magnetic field decreases.
Here, we adopt the magnetic charge representation of the Hamiltonian, Eq.~(\ref{eq:DumbbelHamiltonianMag}), and estimate the critical magnetic field.
At half-magnetization plateau, we have $|Q_p|=2$, for all the plaquettes.
As the magnetic field decreases, we expect the vacuum plaquettes to nucleate.

\begin{figure}[]
\begin{center}
\includegraphics[width=\hsize]{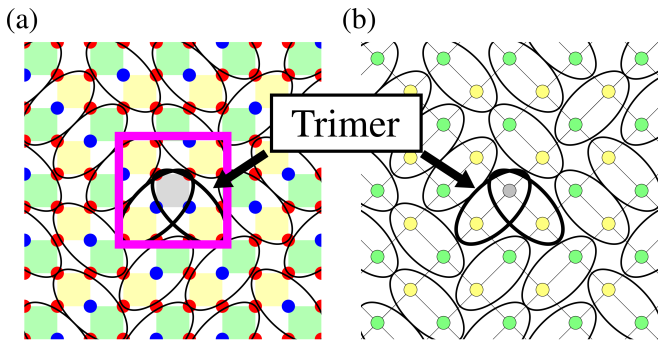}

\end{center}
\caption{(a) A dissociated vacuum plaquette surrounded by charged plaquettes. The yellow and green plaquettes support the magnetic charges, $Q_p=+2$ and $-2$,
as introduced in Fig.~\ref{fig:TopoCharge}.
(b) Dimer representation of the spin configuration shown in (a). The vacuum plaquette can be interpreted as an overlapping part of the two dimers. 
}
\label{fig:trimercovering}
\end{figure}

To examine this process, we take one upward spin and flip it downward, and create a pair of vacuum plaquettes.
The nucleated plaquettes are dissociated, and then they are individually screened by the charged plaquettes [Fig.~\ref{fig:trimercovering} (a)], to maximize the energy gain from the interaction term.
From Eq.~(\ref{eq:hhamiltonian2}), we can estimate the energy increase associated with this process:
\begin{eqnarray}
\Delta E= 4\bigl[(1+2J)S_h - (1+6J)\bigr],
\end{eqnarray}
which leads to the instability at $S_h=\frac{1+6J}{1+2J}$, or
\begin{eqnarray}
h=2+12J,
\end{eqnarray}
below which the dimer phase becomes unstable against the creation of vacuum plaquettes.

\subsubsection{Formation of the trimer CSL}
Next, let us look into the state just below the critical field.
In this state, all the vacuum plaquettes are surrounded by four charged plaquettes with $|Q_p|=2$.
See the structure inside the magenta square in Fig.~\ref{fig:trimercovering}(a).
Since this structure locally optimizes the energy, it is desirable to tile the whole lattice with as many of this local structure as possible.
However, it is not straightforward to obtain the optimal tiling pattern.

\begin{figure}[]
\begin{center}
\includegraphics[width=\hsize]{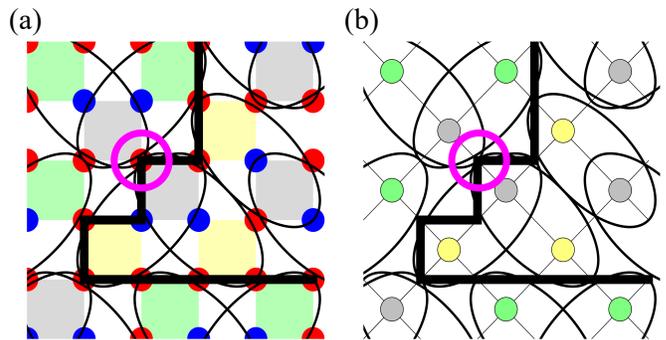}

\end{center}
\caption{(a) An example of prohibited trimer arrangement in the tCSL state. 
The magenta circle represents a touching between two neighboring vacuum plaquettes (``vacuum touching"). 
Bold lines show the boundary between clusters $D_+$ and $D_-$.
The yellow, gray, and green plaquettes support the magnetic charges, $Q_p=+2, 0$ and $-2$, as introduced in Fig.~\ref{fig:TopoCharge}.
(b) Dimer representation of the spin configuration shown in (a).
}
\label{fig:prohibited_trimer}
\end{figure}

To gain an insight, let us adopt a dimer representation we have introduced in Sec.~\ref{sec:Mprocess_negativeJ}, and place a dimer on each downward spin [Fig.~\ref{fig:trimercovering}(b)].
From this viewpoint, a vacuum plaquette can be regarded as an overlapping part of two dimers.
Then the tiling problem can be interpreted as fully packing the dual square lattice with overlapping dimers, or ``trimers",
under the condition that the overlapping part (i.e., a vacuum plaquette) does not neighbor with each other 
(for an example of prohibited structure, see Fig.~\ref{fig:prohibited_trimer}).
The packing with trimers is in sharp contrast to the dimer covering problem that appears on the half-magnetization plateau.
As we numerically estimate in the next subsection, the number of possible trimer configurations increases macroscopically as system size, 
suggesting that the state forms a CSL.
Accordingly, we name this CSL as ``trimer classical spin liquid" (tCSL) after its structure.

\subsubsection{Characterization of trimer CSL}
In this section, we will look into the detailed character of tCSL.
Firstly, the formation of tCSL is associated with the magnetization plateau.
The picture of trimer covering immediately leads to the magnetization value $M=1/3$ per spin, since one out of the three plaquettes forming a trimer has the total spin $S_p=0$, 
while the other two have $S_p=2$. Comparing with the fully polarized ferromagnet, where $S_p=4$ for all the plaquettes, we can easily obtain $M=(0+2+2)/(4\cdot3)=1/3$.

Secondly, what is the ground state energy of tCSL?
To see this, again from the structure of a trimer, the number of vacuum plaquettes is $N_0=N_p/3$, while that of charged plaquette is $N_2=2N_p/3$.
Since each vacuum plaquette contacts with four charged plaquettes, the number of bonds connecting the vacuum and charged plaquettes is $4N_0=\frac{4}{3}N_p$.
Since the total number of bonds is $2N_p$, and no bonds connect two vacuum plaquettes, the number of bonds connecting the two charged plaquettes is $2N_p-\frac{4}{3}N_p=\frac{2}{3}N_p$. Inputting all these into Eq.~(\ref{eq:DumbbelHamiltonianMag}), we obtain the energy of tCSL:
\begin{eqnarray}
E_{\rm tCSL}=\frac{2}{3}(3J-h-1)N_p.
\label{eq:energy_tCSL}
\end{eqnarray}

\begin{figure}[]
\begin{center}
\includegraphics[width=\hsize]{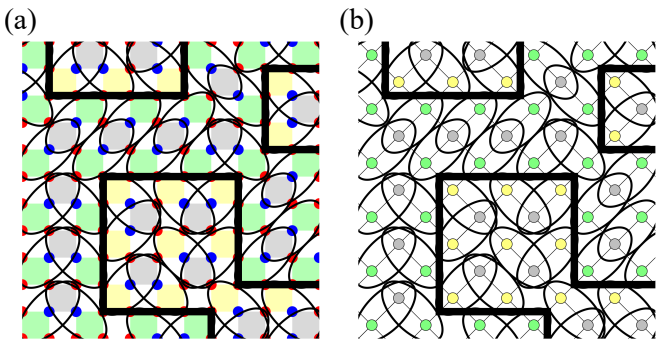}
\end{center}
\caption{
(a) An example of spin configuration in the tCSL state. Bold lines show the boundary between clusters $D_+$ and $D_-$. The yellow, gray, and green plaquettes support the magnetic charge, $Q_p=+2, 0$ and $-2$, as introduced in Fig.~\ref{fig:TopoCharge}.
(b) The trimer covering representation corresponding to the configuration shown in (a).
}
\label{fig:trimer_cluster}
\end{figure}

Thirdly, how large is the ground state degeneracy?
To gain an insight into the origin of degeneracy, let us look at one typical trimer configuration in Fig.~\ref{fig:trimer_cluster}.
Looking at this configuration, one can find a system is divided into large-scale charge clusters.
In a positive (negative) charge cluster, the charged plaquettes are placed only on A (B) sublattice, and the vacuum plaquettes are placed on B (A) sublattices.
Inside one cluster, one finds a staggering pattern of charged and vacuum plaquettes.
Only at the cluster boundaries, the charged plaquettes neighbor with each other.
Meanwhile the vacuum plaquettes are never adjacent to each other.

This cluster structure implies there are two origins for the ground state degeneracy: the contribution from cluster placements,
and the internal spin configuration within a cluster. The existence of these two types of configurational entropy is in common with the hexamer CSL
found in the cousin system of $J_1$-$J_2$-$J_3$ Ising model on the Kagome lattice, where a novel value of residual entropy was found, as well as
a characteristic spin correlation with iconic half-moon pattern in magnetic structure factor\cite{PhysRevLett.119.077207}.

Here, in order to obtain the precise value of residual entropy, we resort to a transfer matrix method, 
by adopting a finite strip of checkerboard lattice with variable widths $L$, which is the number of plaquettes in the column of the strip (see Fig.~\ref{fig:interaction}), up to 7. 
The detail of transfer matrix method is summarized in Appendix C.
We listed the values of residual entropy $\mathcal{S}_0$ for each width on Table~\ref{table:residualentropy}.
While the obtained value still fluctuates around $L\sim7$, they are well within the window of $\mathcal{S}_0=0.136\pm0.04$, 
giving the evidence for the existence of finite residual entropy.
The value $\mathcal{S}_0\simeq0.136$ is smaller compared with the zero-field value, $\mathcal{S}_0^{\rm SI}=0.21556643$, but comparable to that of the dimer phase, $\mathcal{S}_0^{\rm hmp}=0.14578045$. 

Previously, the configurational entropy was obtained for a trimer covering problem in several contexts\cite{frobose1996orientational,ghosh2007random,lee2017resonating,dong20183}.
In Ref.~\onlinecite{frobose1996orientational}, only the angular trimers are considered.
The obtained value is, if translated into the current context, $\mathcal{S}_0^{\rm ang}\sim0.13846575$ per spin,
comparable to our value of $\mathcal{S}_0$.
At first sight, it may seem strange that the limitation to angular trimers do not make difference in larger residual entropy.
However, their model allows the touching of vacuum plaquette in our context, which seems to compensate for the entropy reduction due to the limitation of the type of trimers.
In Ref.~\onlinecite{ghosh2007random}, on the other hand, the authors only discuss the covering by the line trimers, and they obtained  $\mathcal{S}_0^{\rm line}\sim0.07926$.
The covering problems by all species of trimers are considered in Ref.~\onlinecite{dong20183}. They obtained $\mathcal{S}_0^{\rm tot}\sim0.20597$.
The difference between $\mathcal{S}_0^{\rm tot}$ and our $\mathcal{S}_0$ is attributed to the prohibition of vacuum touching.

\begin{table}[htbp]
\begin{center}
  \begin{tabular}{ | c | c | }    
    \hline   
    Width: $L$ & Residual entropy\\ \hline\hline
    2 & 0.14931329 \\ \hline
    3 & 0.15296086 \\ \hline
    4 & 0.13473800 \\ \hline
    5 & 0.13851784 \\ \hline
    6 & 0.13915249 \\ \hline
    7 & 0.13313547 \\   
    \hline
  \end{tabular}
\caption{The width $L$ dependence of residual entropy per site, $\mathcal{S}_0$. } 
\label{table:residualentropy} 
\end{center}
\end{table}

The finite residual entropy qualifies this phase as CSL.
This trimer CSL is similar to the classical spin liquids found in kagome~\cite{PhysRevLett.119.077207} and pyrochlore~\cite{PhysRevB.94.104416} 
lattices previously in a sense that the same-charge attraction leads to these CSL's. In these systems, one spin is fractionalized into a pair of magnetic charges, and the fractionalized charges make recombination into a new phase that cannot be easily inferred from the original spin degrees of freedom.
 
While the trimer CSL has some similarities to the previous CSL's as mentioned above, 
it shows a unique feature in its structural rule: the trimer covering.
To the knowledge of the authors, the trimer model does not belong to known integrable models, 
implying a possibility that this trimer CSL may serve as a precursor of a new type of QSL.

In this light, it is interesting to introduce quantum fluctuation to this system, in the form of, e.g., quantum transverse or exchange interactions.
If QSL is actually realized on the basis of this trimer covering state, we expect it to show an anomalous magnetic charge correlation. 
In a trimer CSL, magnetic charges tend to make clusters, as mentioned above. If this tendency survives after the superposition of trimer configurations, 
it will present strong short-range charge correlation, quite distinct from the previously known class of QSL's.

\subsection{Rigorous argument based on Gauss' law}
\label{sec:tCSL_fromGausslaw}
In the previous subsection, we derived the structure of tCSL in an intuitive way, from the instability analysis of half-magnetization plateau.
Here, we rigorously show that the tCSL state gives the ground state, on the basis of Gauss' law we introduced as Eq.~(\ref{eq:gausslaw}).
To this aim, we rewrite the Hamiltonian Eq.~(\ref{eq:DumbbelHamiltonianMag}) in a form,
\begin{align}
\mathcal{H} &= \left(\frac{1}{2}-J\right)\sum_p(Q_p-\eta_pS_h)^2\nonumber\\
&- J\sum_{\langle p,q\rangle}(Q_p-\eta_pS_h)(Q_q-\eta_qS_h) + C_3,
\label{eq:DumbbelHamiltonianMag_modify}
\end{align}
where
\begin{eqnarray}
C_3 =\frac{N_p}{8}\bigl[16J-16-\frac{h^2}{1+2J}\bigr].
\label{eq:DumbbelHamiltonianMag_const}
\end{eqnarray}
Here, we limit ourselves to small positive $J$, and focus on the magnetic field just below the half-magnetization plateau, i.e., $h\sim2$.
In the range of magnetic field under consideration, we can safely assume that all the plaquettes are occupied by 2-up 2-down or 3-up 1-down spin configuration,
namely, $|Q_p|=0$ or $2$, and $Q_p = +2$ $(-2)$ exists only at the $A$ ($B$) sublattice.
Under this assumption, let us define the number of charges with $|Q_p|=0\ (2)$ as $N_0$ ($N_2$), and the number of contacts between the plaquettes with $|Q_p|=0$ and $|Q_q|=2$ as $n_{20}$. Similarly, we define $n_{00}$ ($n_{22}$) as the number of contacts between two plaquettes with charge $0$ ($2$).
With these quantities, we can express the Hamiltonian, Eq.~(\ref{eq:DumbbelHamiltonianMag_modify}) as
\begin{align}
\mathcal{H}&=\left(\frac{1}{2}-J\right)S_h^2N_0+\left(\frac{1}{2}-J\right)(S_h-2)^2N_2 \nonumber \\ 
&+JS_h(S_h-2)n_{20}+JS_h^2n_{00}+J(S_h-2)^2n_{22}.
\label{eq:Ham_charge_contact}
\end{align}

The variables in this Hamiltonian are subject to several geometrical constraints.
Firstly, since each plaquette has four contacts with neighboring plaquettes on a dual square lattice, we have
\begin{eqnarray}
4N_0 = 2n_{00} + n_{20},
\label{eq:Geometrical_equality_N0}
\end{eqnarray}
and
\begin{eqnarray}
4N_2 = 2n_{22} + n_{20}.
\label{eq:Geometrical_equality_N2}
\end{eqnarray}
Here, the factor 2 before $n_{00}$ and $n_{22}$ correct the double counting.
Secondly, in the absence of double charges ($|Q_p|=4$), the sum of $N_0$ and $N_2$ is equal to the total number of plaquettes:
\begin{eqnarray}
N_0 + N_2 = N_p.
\label{eq:Geometrical_equality_Np}
\end{eqnarray}
Combining Eqs.~(\ref{eq:Geometrical_equality_N0})-(\ref{eq:Geometrical_equality_Np}), the contact numbers satisfy
\begin{eqnarray}
n_{00} + n_{22} + n_{20} = 2N_p.
\label{eq:contact_sumrule}
\end{eqnarray}
 
With Eqs.~(\ref{eq:Geometrical_equality_N0}) and (\ref{eq:Geometrical_equality_N2}), we can eliminate $N_0$ and $N_2$ from the Hamiltonian (\ref{eq:Ham_charge_contact}), and obtain
\begin{align}
\mathcal{H} &= \bigl[\frac{1}{2}(\frac{1}{2}+J)S_h^2\bigr]n_{00}+\bigl[\frac{1}{2}(\frac{1}{2}+J)(S_h-2)^2\bigr]n_{22} \nonumber \\
&+\bigl[\frac{1}{2}(\frac{1}{2}+J)(S_h-1)^2+\frac{1}{4}-\frac{3}{2}J\bigr]n_{20}\nonumber\\
&=a_{00}n_{00}  + a_{22}n_{22}+ a_{20}n_{20}, 
 \label{eq:Ham_contact}
\end{align}
where $a_{00}$, $a_{22}$ and $a_{20}$ are coefficients of  $n_{00}$, $n_{22}$ and $n_{20}$.
Accordingly, the search for the ground state is now reduced to finding the combination of $(n_{00}, n_{22}, n_{20})$ to minimize $\mathcal{H}$,
under the constraint of sum rule, Eq.~(\ref{eq:contact_sumrule}).

\begin{figure}[]
\begin{center}
\includegraphics[width=\hsize]{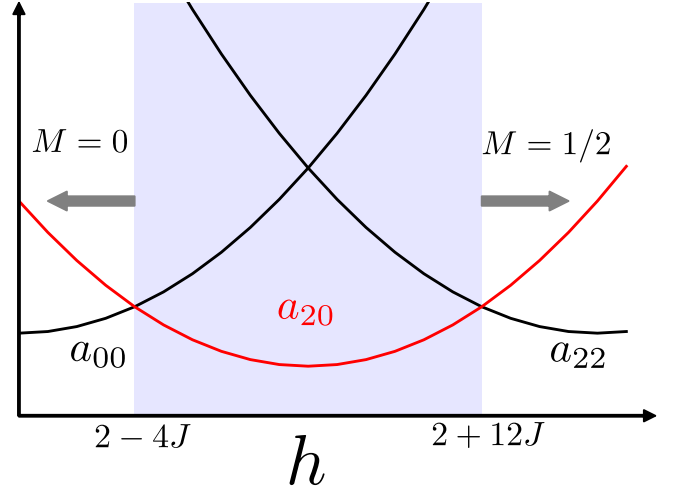}
\end{center}
\caption{
The coefficients of the Hamiltonian, Eq.~(\ref{eq:Ham_contact}). For $2-4J<h<2+12J$, the coefficient of $n_{20}$ is lower than those of $n_{00}$ and $n_{22}$.}
\label{fig:schema_coef}
\end{figure}

The coefficients $a_{00}$, $a_{22}$ and $a_{20}$ are plotted in Fig.~\ref{fig:schema_coef}.
For $h>2+12J$, $a_{22}$ is the smallest. This region corresponds to the half-magnetization plateau, where all the plaquettes are occupied with charges $|Q_p|=2$,
and all the contacts are of $2$-$2$ type, accordingly.

Meanwhile, this plot shows that larger $n_{20}$ is preferable for $2-4J<h<2+12J$.
To satisfy this condition, at first sight, the best strategy seems to put $Q_p=2$ on all the plaquettes of A sublattice, while $Q_p=0$ on B sublattice, to make all the contacts to be of 2-0 type.
However, this charge configuration obviously violates the Gauss' law, Eq.~(\ref{eq:gausslaw}). 

To find the optimal charge configuration under the Gauss' constraint, let us define a positive (negative) charge cluster $D_+$ ($D_-$), 
as a maximal set of plaquettes with $Q_p=2$ placed on the A(B)-sublattice and those with $Q_p=0$ placed on the B(A)-sublattice [Fig.~\ref{fig:trimer_cluster} (a)]. Namely,
\begin{align}
\mathrm{cluster\:}D_+:Q_p=
\{
\begin{array}{ll}
+2 & (\mathrm{for}\; A\; \mathrm{sub.}) \\
0 & (\mathrm{for}\; B\; \mathrm{sub.})
\end{array}, \\
\mathrm{cluster\:}D_-:Q_p=
\{
\begin{array}{ll}
0 & (\mathrm{for}\; A\; \mathrm{sub.}) \\
-2 & (\mathrm{for}\; B\; \mathrm{sub.})
\end{array}.
\end{align}
With these definitions, clusters $D_+$ and $D_-$ always touch with each other through the 0-0 or 2-2 contacts.
To see this, suppose a cluster $D_+$ has a boundary plaquette that belongs to A (B) sublattice, then it must have charge $|Q_p|=2$ (0). 
This plaquette neighbors with a plaquette of $D_-$ on the B (A) sublattice, which has charge $|Q_p|=2$ (0).

Now, let us apply the Gauss' law to a cluster, $D_{\alpha}$ of either type.
We assume that the number of charged plaquettes inside $D_{\alpha}$ to be $N_2^{\alpha}$, and define the number of 0-0 and 2-2 contacts with neighboring clusters to be 
$n_{00}^{\alpha}$ and $n_{22}^{\alpha}$, respectively.
The Gauss' inequality, Eq.~(\ref{eq:gaussinequality}) leads to
\begin{eqnarray}
2N_2^{\alpha}=|\sum_{p\in D_{\alpha}}Q_p|\leq\sum_{i\in\partial D_{\alpha}}|\eta_{p}\sigma^z_i|=n_{22}^{\alpha}+n_{00}^{\alpha}.
\label{eq:application_Gaussinequality}
\end{eqnarray}
By summing over all the clusters in the system, we obtain
\begin{eqnarray}
2N_2\leq 2(n_{00}+n_{22}).
\label{eq:Gauss_constraint}
\end{eqnarray}
Note that the factor 2 of the right-hand side comes from the double counting of bonds in the summation over clusters.
Combining Eqs.~(\ref{eq:Geometrical_equality_N2}) and (\ref{eq:Gauss_constraint}), we obtain
\begin{eqnarray}
n_{20}\leq 4n_{00} + 2n_{22},
\label{eq:Gauss_contact_inequality}
\end{eqnarray}
 
Now, considering the relative magnitudes of coefficients depicted in Fig.~\ref{fig:schema_coef}, at $h\lesssim2+12J$, Eq.~(\ref{eq:Gauss_contact_inequality}) results in the optimal solution to be $n_{00}=0$ and $n_{20}=2n_{22}$.
If a certain configuration satisfies this condition, it gives one of the ground states. In fact, this condition is equivalent to the trimer covering we discussed in the previous subsection.
By inserting this condition into Hamiltonian~(\ref{eq:Ham_charge_contact}) with the constant term $C_3$ given by Eq.~(\ref{eq:DumbbelHamiltonianMag_const}), we obtain 
\begin{eqnarray}
E_{\rm tCSL}=\frac{2}{3}(3J-h-1)N_p,
\label{eq:GSenergy}
\end{eqnarray}
which exactly corresponds to what we obtained from the trimer covering picture: Eq.~(\ref{eq:energy_tCSL}).
This means that the tCSL states give the ground state.

Conversely, it is possible to show that any member of the ground state manifold can be expressed by the trimer covering, i.e., the ground state is composed only of the tCSL states.
To prove this, it is enough to show that if $n_{00}=0$ and $n_{20}=2n_{22}$ are satisfied, the corresponding charge configuration can be expressed in terms of the trimer covering.

To see this, suppose a cluster $D_{\alpha}\in D_+$. 
To satisfy the former condition, one needs $n_{00}^{\alpha}=0$, and in addition to that, the equality must hold in the Gauss' inequality, Eq.~(\ref{eq:application_Gaussinequality}):
\begin{eqnarray}
2N_2^{\alpha}=n_{22}^{\alpha}=\sum_{i\in\partial D_{\alpha}}\sigma_i^z,
\end{eqnarray}
This equality requires the boundary spins to satisfy $\sigma_i^z=+1$, and the cluster $D_{\alpha}$ to neighbor with other clusters with charged plaquette.
Accordingly, within the cluster $D_{\alpha}$, each charged plaquette shares its only down spin with its neighboring vacuum plaquette.
Moreover, each vacuum plaquette shares its two down spins with two of its neighboring charged plaquettes.
Consequently, if one places a dimer on each down spin, all the charged plaquettes are covered with one dimer, and all the vacuum plaquettes are covered with two dimers,
resulting in a trimer covering. The same argument holds for a cluster $D_{\alpha}\in D_-$.

\section{Full magnetization process \label{sec:magnetization}}
\begin{figure}[]
\begin{center}
\includegraphics[width=\hsize]{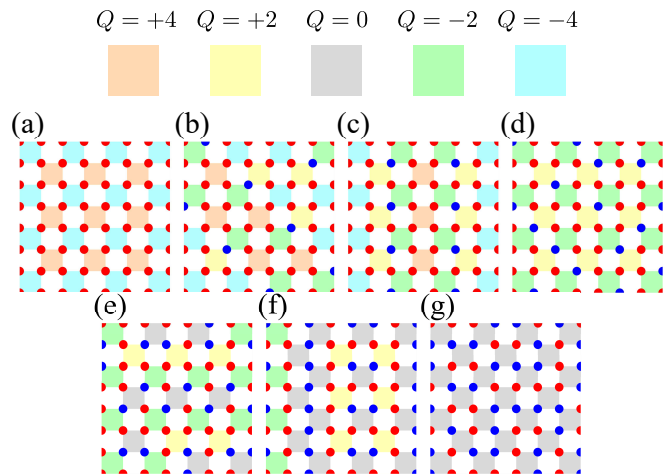}
\end{center}
\caption{
The schematic pictures of spin configurations at each phase, corresponding to the magnetization plateaus at (a) $M=1$, (b) $M=3/4$, (c) $M=2/3$, (d) $M=1/2$, (e) $M=1/3$, 
(f) $M=1/5$, and (g) $M=0$.  
}
\label{fig:M_high}
\end{figure}

\begin{figure}[h]
\begin{center}
\includegraphics[width=\hsize]{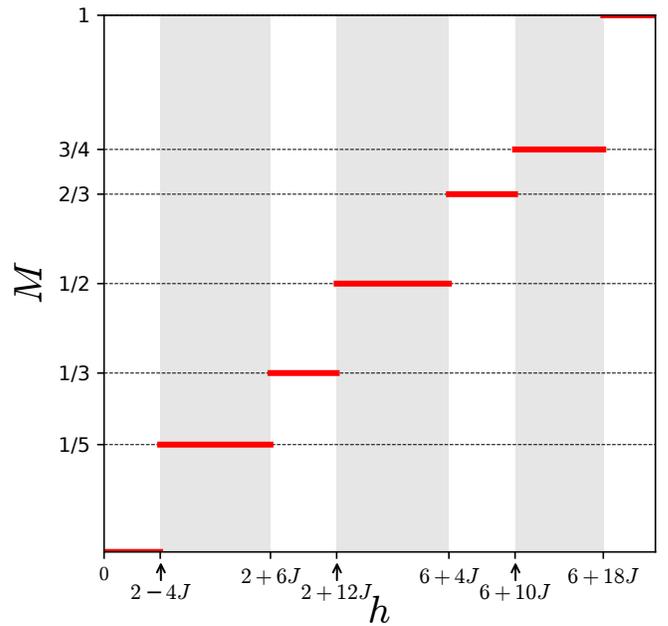}
\end{center}
\caption{The magnetization process for small $J>0$. 
There are seven phases that have $M=0, 1/5, 1/3, 1/2, 2/3, 3/4$ and $1$. The corresponding transition points are $h=2-4J, 2+6J, 2+12J, 6+4J, 6+10J$ and $6+18J$. 
Shaded areas are guides for the eyes.}
\label{fig:magnetization_process}
\end{figure}

In this section, we will address the rest of magnetic phase diagram shown in Fig.~\ref{fig:PhaseDiagram} for $J>0$.
We limit ourselves to the region of small $J$, again. 
We show the schematic picture of spin configurations at each phase in Fig.~\ref{fig:M_high}, and the magnetization process in Fig.~\ref{fig:magnetization_process}.

\subsection{Instability of high-field phase}
We start with the instability of the high-field fully-polarized phase [Fig.~\ref{fig:M_high}(a)], with decreasing $h$.
In the fully-polarized phase, all the plaquettes have total spins $S_p=4$.
A single spin flip makes an adjacent pair of plaquettes with $S_p=2$.
The energy increase accompanying this process is
\begin{eqnarray}
\Delta E= 4\bigl[(1+2J)(S_h-3)-3J\bigr].
\end{eqnarray}
This instability occurs for $\Delta E<0$, i.e.
\begin{eqnarray}
h<6+18J.
\end{eqnarray}
Below this boundary line, the system tries to maximize the number of neighboring pairs of plaquettes with $S_p=4$ and $S_p=2$.
Consequently, we obtain the 4-plaquette ordering as shown in Fig.~\ref{fig:M_high}(b). 
This ordered phase corresponds to the $M=3/4$ plateau.

\subsection{High-field instability of dimer phase}
In the previous section, we addressed the low-field instability of half-magnetization plateau into tCSL.
Here, we address the instability with an increasing magnetic field.
In the half-magnetization plateau, all the plaquettes have $S_p=2$ uniformly.
As the magnetic field increases, we expect a nucleation of double charge, $S_p=4$.
The nucleation process takes qualitatively different form from the nucleation of $S_p=2$ plaquette from the fully-polarized phase.

In the fully-polarized phase, the nucleated $S_p=2$ plaquettes are always paired.
In contrast, in the half-magnetization plateau, the nucleation of $S_p=4$ plaquettes occurs in pair, but they
can be dissociated from each other. 
This is a sort of fractionalization, which reflects the deconfining nature of the dimer phase at the half-magnetization plateau.

As a result, the most economical excitation is a single plaquette of $S_p=4$ surrounded by $S_p=2$ plaquettes.
This object costs energy
\begin{eqnarray}
\Delta E=4\bigl[(1+2J)(3-S_h)-4J\bigr],
\end{eqnarray}
which becomes negative, if
\begin{eqnarray}
h>6+4J.
\end{eqnarray}
Above this field, the spin configuration, as shown in Fig.~\ref{fig:M_high}(c), is stabilized.

If we stand on the dimer picture by placing a dimer on a down spin, the resultant state consists of an assembly of monomers screened by dimers.
It follows naturally that the corresponding state makes a $M=2/3$ plateau.
Most simply, three-plaquette ordering realizes this state [Fig.~\ref{fig:M_high}(c)], however, introduction of ``stacking fault" does not increase the energy.
Accordingly, this state shows a semi-macroscopic degeneracy of the order of $2^{L}$.

Obviously, this $M=2/3$ plateau cannot be continuously connected to the 4-plaquette ordering of $M=3/4$ plateau, just below the saturated phase.
The transition between the two states occurs at $h=6+10J$, from a simple comparison of the energies.

\subsection{Instability of square ice}
Finally, we address the instability of square ice state, as increasing $h$ from 0.
To this aim, rather than the dimer representation, we resort to Gauss' inequality, following the argument in Sec.~\ref{sec:tCSL_fromGausslaw}.

We start with Hamiltonian~(\ref{eq:Ham_contact}).
From their coefficients as shown in Fig.~\ref{fig:schema_coef}, it follows that $n_{22}$ should be suppressed for smaller magnetic field.
Meanwhile larger $n_{20}$ is preferable for $h>2-4J$.

Given that $n_{22}=0$, a set of geometrical equations, Eqs.~(\ref{eq:Geometrical_equality_N0}), (\ref{eq:Geometrical_equality_N2}), and (\ref{eq:contact_sumrule})
lead to
\begin{align}
4N_0=2n_{00}+n_{20}, \\
4N_2=n_{20}, \\
n_{00}+n_{20}=2N_p.
\label{eq:Geometrical_constraint_lowfield}
\end{align}
The optimal value of $n_{20}$ is constrained by the Gauss' inequality, Eq.~(\ref{eq:Gauss_constraint}) as
\begin{align}
2N_2\leq 2n_{00},
\end{align}
which leads to the inequality,
\begin{align}
n_{20}\leq 4n_{00},
\label{eq:Gaussinequality_lowfield}
\end{align}
combined with the geometrical constraint, Eq.~(\ref{eq:Geometrical_constraint_lowfield}).

The optimal value of $n_{20}$ corresponds to the equality of Eq.~(\ref{eq:Gaussinequality_lowfield}), which leads to
\begin{align}
n_{00}=\frac{2}{5}N_p,\ \ n_{20}=\frac{8}{5}N_p,
\end{align}
and
\begin{align}
N_0=\frac{3}{5}N_p,\ \ N_2=\frac{2}{5}N_p.
\end{align}
This state corresponds to the $M=1/5$ plateau, and we show the configuration in Fig.~\ref{fig:M_high}(f). 

\section{Summary \label{sec:summary}}
We have studied a problem of interacting fractional charges, taking the $J_1$-$J_2$-$J_3$ Ising model on a checkerboard lattice under the magnetic field.
We focused on the case, where the Hamiltonian can be written in terms of the charge degrees of freedom.
In particular, at small positive $J$, we found that the half-magnetization plateau destabilizes into a classical spin liquid state, as decreasing the magnetic field.
The resultant CSL is expressed as an assembly of trimer covering of the dual square lattice, and we called this state the tCSL.

The tCSL state corresponds to the $1/3$ magnetization plateau, and has macroscopic ground state degeneracy, which is characterized by a novel value of residual entropy.
In contrast to dimer covering, which is ubiquitous in a broad area of physics including statistical mechanics and condensed matter physics,
the notion of trimer covering rarely appears.
It is surprising that such elusive states can be obtained, by starting from a simple local Hamiltonian considered here.
As turning on a quantum fluctuation, a novel quantum spin liquid state may further be stabilized based on a well-defined microscopic model. 

Moreover, we showed the interactions among magnetic charges lead to a variety of magnetization plateaus in the applied magnetic field, reflecting the rich screening processes of dimer-monomer mixtures. Nontrivial magnetization processes are observed for a number of frustrated magnetic systems.
In this regard, this work shows that the picture of interacting fractional charges gives a new viewpoint to the formation of magnetization plateaus.

\acknowledgements
This work was supported by the JSPS KAKENHI (Nos. JP15H05852, JP16H04026, and JP17H06138), MEXT, Japan. Part of
numerical calculations were carried out on the Supercomputer Center at Institute for Solid State Physics, University of Tokyo. 
K. T. was supported by the Japan Society for the Promotion of Science through the Program for Leading Graduate Schools (MERIT).

\appendix
\section{Mapping to charge Hamiltonian}
In this appendix, we derive the charge representation of Hamiltonian Eq.~(\ref{eq:DumbbelHamiltonianMag}) 
from the original Hamiltonian Eq.~(\ref{eq:hamiltonian}).
First, we rewrite Eq.~(\ref{eq:hamiltonian}) 
with $(J_1,J_2,J_3) = (1,2J,J)$
as 
\begin{figure}[]
\begin{center}
\includegraphics[width=0.7\hsize]{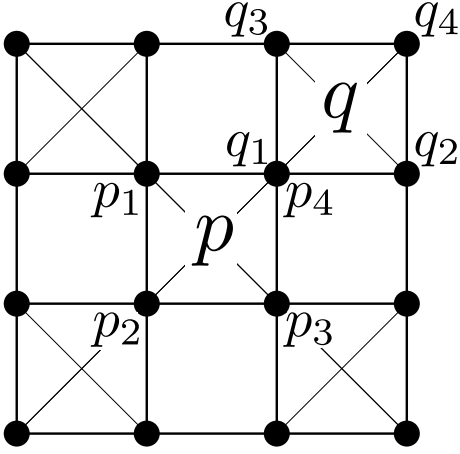}
\end{center}
\caption{The definition of charges $p, q$ and their components spins $p_i$ and $q_i$. We define sites $[p,q]$ as be shared by two charges $p$ and $q$, therefore $[p,q]=p_4=q_1$.}
\label{fig:apeA}
\end{figure}

\begin{align}
\mathcal{H}=&\frac{1}{2}\sum_p(\sigma_{p_1}^z+\sigma_{p_2}^z+\sigma_{p_3}^z+\sigma_{p_4}^z)^2-2N_p \nonumber \\
&+J\sum_{\langle p,q \rangle}(\sigma_{p_1}^z+\sigma_{p_2}^z+\sigma_{p_3}^z+\sigma_{p_4}^z-\sigma_{[p,q]}^z)\nonumber \\
&\times(\sigma_{q_1}^z+\sigma_{q_2}^z+\sigma_{q_3}^z+\sigma_{q_4}^z-\sigma_{[p,q]}^z)\nonumber \\
&-\frac{h}{2}\sum_p(\sigma_{p_1}^z+\sigma_{p_2}^z+\sigma_{p_3}^z+\sigma_{p_4}^z) \label{eq:hamiltonianApeA1},
\end{align}
where we name the spins on the neighboring plaquettes, $p$ and $q$, as $p_1$-$p_4$ and $q_1$-$q_4$ respectively, 
as shown in Fig.~\ref{fig:apeA}. 
The symbol $[p,q]$ stands for the spin shared by $p$ and $q$. For instance, $[p,q]=p_4=q_1$ in Fig.~\ref{fig:apeA}.
The first and the second term come from the nearest-neighbor interactions 
while the third term comes from the second-neighbor and the third-neighbor interactions 
which connect spins on neighboring plaquettes. 
The fourth term comes from magnetic field.
We transform Eq.~(\ref{eq:hamiltonianApeA1}) by introducing the total spin of the plaquettes $S_p=\sum_{i\in p}\sigma^{z}_i$:
\begin{align}
\frac{1}{2}\sum_p(\sigma_{p_1}^z+\sigma_{p_2}^z+\sigma_{p_3}^z+\sigma_{p_4}^z)^2 =& \frac{1}{2}\sum_p S_p^2, \\
\frac{h}{2}\sum_p(\sigma_{p_1}^z+\sigma_{p_2}^z+\sigma_{p_3}^z+\sigma_{p_4}^z) =&\frac{h}{2}\sum_pS_p,
\end{align}
and
\begin{align}
&J\sum_{\langle p,q \rangle}(\sigma_{p_1}^z+\sigma_{p_2}^z+\sigma_{p_3}^z+\sigma_{p_4}^z-\sigma_{[p,q]}^z)\nonumber \\
&\times(\sigma_{q_1}^z+\sigma_{q_2}^z+\sigma_{q_3}^z+\sigma_{q_4}^z-\sigma_{[p,q]}^z)\nonumber \\
=&J\sum_{\langle p,q \rangle}(S_p-\sigma_{[p,q]}^z)(S_q-\sigma_{[p,q]}^z)\nonumber \\
=&J\sum_{\langle p,q \rangle}(S_pS_q-S_p\sigma_{[p,q]}^z-S_q\sigma_{[p,q]}^z)+J\sum_{\langle p,q \rangle}1\nonumber \\
=&J\sum_{\langle p,q \rangle}S_pS_q-J\sum_{p}S_p\sum_{q\in (\mathrm{n.n.\: of\:} p)}\sigma_{[p,q]}^z+2JN_p \nonumber \\
=&J\sum_{\langle p,q \rangle}S_pS_q-J\sum_{p}S_p^2 +2JN_p.
\end{align}
By introducing charge representation $Q_p=\eta_p S_p$, finally we obtain Eq.~(\ref{eq:DumbbelHamiltonianMag}):
\begin{align}
\mathcal{H}=\left( \frac{1}{2} -J \right) \sum_{p}Q_p^2-J\sum_{\langle p,q \rangle}Q_pQ_q 
- \frac{h}{2}\sum_p\eta_pQ_p + C_1, 
\end{align}
where $C_1 = 2(J-1)N_p$. 

\section{Ordered phase at $h=0$ for $J>1/4$.}
\begin{figure}[]
\begin{center}
\includegraphics[width=\hsize]{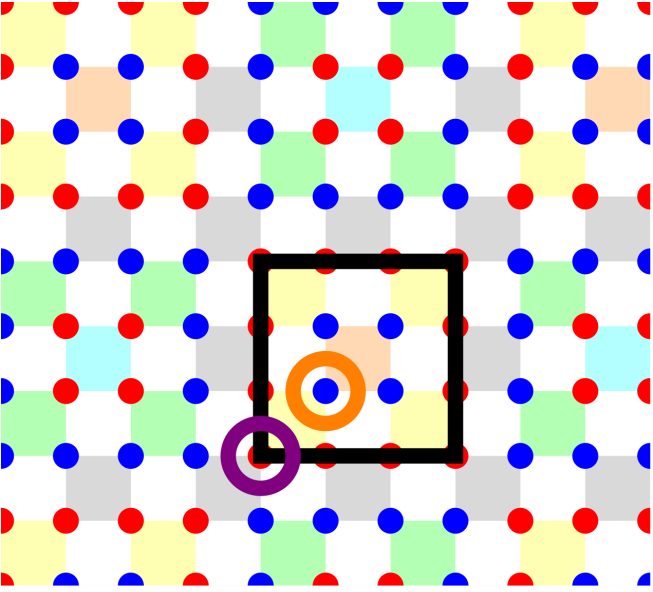}
\end{center}
\caption{The schematic picture of spin configurations in the absence of magnetic field for $J>1/4$ [same as Fig.~\ref{fig:TopoCharge}(d)]. 
Bold line represents one of the same-sign-charge cluster $\mathcal{D}$. 
The site enclosed by orange (purple) circle is an example of a site counted as $n_{i:(2.4)}^{(\mathcal{D})}$ ($n_{b:(2.4)}^{(\mathcal{D})}$). }
\label{fig:apeC}
\end{figure}

In this appendix, 
we derive the phase boundary for $h=0$ and $J>0$, 
by evaluating the ground-state energy using the constraint of the Gauss' law and some geometrical identities.
We show the Coulomb phase survives up to $J=1/4$, and the ordered phase in Fig.~\ref{fig:apeC} realizes for $J>1/4$.

The outline of the energy evaluation is as follows. 
First, we decompose an arbitrary 
charge configuration into assembly of maximally connected same-sign-charge clusters and vacuum plaquettes.
Here the maximally connected same-sign-charge cluster means that 
it must be connected to opposite-sign-charges or vacuum plaquettes at the boundary.
By doing so, the total energy is written by the sum of the intra-cluster energy and the inter-cluster energy.
Then, we optimize the energy under the geometrical constraint and the Gauss' law, 
to obtain the ground-state charge configuration.

To begin with, let us address an essential property of the same-sign-charge cluster (labeled as $\mathcal{D}$), 
which originates from the Gauss' law. 
Namely, applying the triangle inequality Eq.~(\ref{eq:gaussinequality}) to $\mathcal{D}$,
we have
\begin{align}
2 N_2^{(\mathcal{D})} + 4N_4^{(\mathcal{D})} \leq n_b^{(\mathcal{D})}, 
\label{eq:gauss_cluster}
\end{align}
where 
$N_2^{(\mathcal{D})}$ ($N_4^{(\mathcal{D})}$) is a number of $|Q_p|=2$ ($|Q_p|=4$) plaquettes in $\mathcal{D}$, and
$n_b^{(\mathcal{D})}$ denotes the number of boundary sites of $\mathcal{D}$.

Having this at hand, we now evaluate the ground state energy. To do this, we introduce the following variables. 
(1) $n_{i:(\ell.\ell^\prime)}^{(\mathcal{D})}$: a number of inner sites of $\mathcal{D}$-clusters which is contact points of $|Q_p| = \ell$ and $|Q_p| = \ell^\prime$.
(2) $n_{b:(\ell.\ell^\prime)}^{(\mathcal{D})}$: a number of boundary sites of $\mathcal{D}$-clusters
at which a plaquette with $|Q_p| = \ell$ {\it inside} the cluster confronts another plaquette with $|Q_p| = \ell^\prime$ {\it outside} the cluster.
See Fig.~\ref{fig:apeC} for examples of sites counted as $n_{i:(2.4)}^{(\mathcal{D})}$ and $n_{b:(2.4)}^{(\mathcal{D})}$.
We note that 
the total number of inner sites $n_i^{(\mathcal{D})}$ and 
that of boundary sites $n_b^{(\mathcal{D})}$ satisfy
\begin{align}
n_i^{(\mathcal{D})}=&n_{i:(2.2)}^{(\mathcal{D})}+n_{i:(2.4)}^{(\mathcal{D})},  \nonumber \\
n_b^{(\mathcal{D})}=&n_{b:(2.0)}^{(\mathcal{D})}+n_{b:(2.2)}^{(\mathcal{D})}+n_{b:(2.4)}^{(\mathcal{D})} \nonumber \\
&+n_{b:(4.0)}^{(\mathcal{D})}+n_{b:(4.2)}^{(\mathcal{D})}+n_{b:(4.4)}^{(\mathcal{D})}. \label{eq:identity0}
\end{align}
Note that $n_{i:(4.4)}^{(\mathcal{D})} = 0$, since the same-sign charges with $|Q_p|=4$ cannot touch with each other.  
Using these valuables, the total energy can be written as
\begin{align}
E=& \left(\frac{1}{2}-J \right)\left(\sum_{\mathcal{D}} 16N_4^{(\mathcal{D})}+4N_2^{(\mathcal{D})}\right)\nonumber \\
&-J\sum_{\mathcal{D}} \left(4n_{i:(2.2)}^{(\mathcal{D})}+8n_{i:(2.4)}^{(\mathcal{D})} \right) \nonumber \\
&+\frac{J}{2}\sum_{\mathcal{D}} \left(4n_{b:(2.2)}^{(\mathcal{D})}+8n_{b:(2.4)}^{(\mathcal{D})}+8n_{b:(4.2)}^{(\mathcal{D})}+16n_{b:(4.4)}^{(\mathcal{D})} \right). \label{eq:Energy1}
\end{align}
Here, the terms proportional to $n_i^{(\mathcal{D})}$ ($n_b^{(\mathcal{D})}$) correspond to intra-cluster (inter-cluster) energy.
To simplify Eq.~(\ref{eq:Energy1}) by erasing $n_{i:(2.2)}^{(\mathcal{D})}$, 
we substitute the following identities,
\begin{align}
4N^{(\mathcal{D})}&=2n_i^{(\mathcal{D})}+n_b^{(\mathcal{D})}, \label{eq:ND=} \\
4N_4^{(\mathcal{D})}&=n_{i:(2.4)}^{(\mathcal{D})}+n_{b:(4.4)}^{(\mathcal{D})}+n_{b:(4.2)}^{(\mathcal{D})}+n_{b:(4.0)}^{(\mathcal{D})}, \label{eq:identity1}
\end{align}
into Eq.~(\ref{eq:Energy1}), and then we obtain
\begin{align}
E=&\left(2-12J \right) \left(N_p-N^{\mathrm{vac}}\right) +2J\sum_{\mathcal{D}} n^{(\mathcal{D})}_b\nonumber \\
&+\left(\frac{3}{2}-7J \right)\sum_{\mathcal{D}} n_{i:(2.4)}^{(\mathcal{D})} \nonumber \\
&+\sum_{\mathcal{D}}\biggl[2Jn_{b:(2.2)}^{(\mathcal{D})}+4J n_{b:(2.4)}^{(\mathcal{D})} \nonumber \\
&+\left(\frac{3}{2}-3J \right) n_{b:(4.0)}^{(\mathcal{D})}+ \left(\frac{3}{2}+J \right) n_{b:(4.2)}^{(\mathcal{D})} \nonumber \\
&+\left(\frac{3}{2}+5J \right) n_{b:(4.4)}^{(\mathcal{D})}\biggr],\label{eq:Energy2} 
\end{align}
where $N_p=\sum_{\mathcal{D}}N_4^{(\mathcal{D})}+\sum_{\mathcal{D}}N_2^{(\mathcal{D})}+N^{\mathrm{vac}}$ 
is a total number of plaquettes and $N^{\mathrm{vac}}$ is a number of vacuum plaquettes.
Applying the inequality of Eq.~(\ref{eq:gauss_cluster}) to 
the right side of
Eq.~(\ref{eq:Energy2}) and using Eq.~(\ref{eq:identity1}), we obtain
\begin{align}
E\geq&8\left(\frac{1}{4}-J \right) \left(N_p-N^{\mathrm{vac}} \right)+6\left(\frac{1}{4}-J \right)
\sum_{\mathcal{D}} n_{i:(2.4)}^{(\mathcal{D})} \nonumber \\
&+\sum_{\mathcal{D}}\biggl[2Jn_{b:(2.2)}^{(\mathcal{D})}+4J n_{b:(2.4)}^{(\mathcal{D})} \nonumber \\
&+2\left(\frac{3}{4}-J \right)
n_{b:(4.0)}^{(\mathcal{D})}
+\left(\frac{3}{2}+2J \right) n_{b:(4.2)}^{(\mathcal{D})} \nonumber \\
&+ \left(\frac{3}{2} + 6J \right)n_{b:(4.4)}^{(\mathcal{D})}\biggr]. \label{eq:finalform2}
\end{align}
For $0 < J < 1/4$, all the coefficients of $n^{(\mathcal{D})}_{i:(\ell,\ell^\prime)}$, $n^{(\mathcal{D})}_{b:(\ell,\ell^\prime)}$ and $\left(N_p-N^{\mathrm{vac}} \right)$ 
in the right side of Eq.~(\ref{eq:finalform2})
are positive, 
thus the energy minimization is achieved when $n^{(\mathcal{D})}_{i:(\ell,\ell^\prime)} = n^{(\mathcal{D})}_{b:(\ell,\ell^\prime)}= \left(N_p-N^{\mathrm{vac}} \right) = 0$,
which is nothing but the Coulomb phase. 
For $J>1/4$, on the other hand,
the coefficients of $n^{(\mathcal{D})}_{i:(2.4)}$ and $\left(N_p-N^{\mathrm{vac}} \right)$ become negative, indicating the instability of the Coulomb phase. 

How can we derive the ground state for $J>1/4$?
 To do this, let us first derive the upper bound of $n_{i:(2.4)}^{(\mathcal{D})}$.
To this end, we point out that the following identity holds:
\begin{align}
4N_2^{(\mathcal{D})}&=n_{i:(2.4)}^{(\mathcal{D})}+ 2n_{i:(2.2)}^{(\mathcal{D})}+n_{b:(2.0)}^{(\mathcal{D})}+n_{b:(2.2)}^{(\mathcal{D})}+n_{b:(2.4)}^{(\mathcal{D})}.\label{eq:identity2}
\end{align}
Then, combining Eqs.~(\ref{eq:gauss_cluster}), 
(\ref{eq:identity0}), (\ref{eq:identity1}), and (\ref{eq:identity2}), after some of algebra, we obtain
\begin{align}
3n_{i:(2.4)}^{(\mathcal{D})}\leq n_{b:(2.0)}^{(\mathcal{D})}+n_{b:(2.2)}^{(\mathcal{D})}+n_{b:(2.4)}^{(\mathcal{D})}. \label{eq:constraint2}
\end{align}
Combining Eqs.~(\ref{eq:constraint2}) and (\ref{eq:finalform2}), we have
\begin{align}
E\geq&8 \left(\frac{1}{4}-J \right)\biggl[N_p-N^{\mathrm{vac}}\biggr] \nonumber \\
&+\sum_{\mathcal{D}}\biggl[(\frac{1}{2}-2J) n_{b:(2.0)}^{(\mathcal{D})} + \frac{1}{2}n_{b:(2.2)}^{(\mathcal{D})}+ \left(\frac{1}{2}+2J \right)
n_{b:(2.4)}^{(\mathcal{D})}\nonumber \\
&+2 \left(\frac{3}{4}-J \right)n_{b:(4.0)}^{(\mathcal{D})}+ \left(\frac{3}{2}+2J \right)n_{b:(4.2)}^{(\mathcal{D})}\nonumber \\
&+ \left(\frac{3}{2}+6J \right)n_{b:(4.4)}^{(\mathcal{D})}\biggr] . \label{eq:finalform3}
\end{align}
Further, we can derive the lower bound of $N^{\mathrm{vac}}$ as
\begin{align}
4N^{\mathrm{vac}} =& 2 N_{i:(0.0)}
+ \sum_{\mathcal{D}}n_{b:(2.0)}^{(\mathcal{D})} + \sum_{\mathcal{D}}n_{b:(4.0)}^{(\mathcal{D})} \notag \\
\geq & \sum_{\mathcal{D}}n_{b:(2.0)}^{(\mathcal{D})}+\sum_{\mathcal{D}}n_{b:(4.0)}^{(\mathcal{D})}, \label{eq:constraint3}
\end{align}
where $N_{i:(0.0)}$ is the number of contacts between vacuum plaquettes. 
Combining Eqs.~(\ref{eq:constraint3}) and (\ref{eq:finalform3}), we obtain
\begin{align}
E \geq & 8 \left(\frac{1}{4}-J \right) N_p + \sum_{\mathcal{D}  }\biggl[\frac{1}{2}n_{b:(2.2)}^{(\mathcal{D})} + \left(\frac{1}{2}+2J \right) n_{b:(2.4)}^{(\mathcal{D})}\nonumber \\
&+n_{b:(4.0)}^{(\mathcal{D})}+ \left(\frac{3}{2}+2J \right) n_{b:(4.2)}^{(\mathcal{D})}
+\left(\frac{3}{2}+6J \right)n_{b:(4.4)}^{(\mathcal{D})}\biggr] . \label{eq:finalform4}
\end{align}
Now, we see from Eq.~(\ref{eq:finalform4}) that the energy minimization is achieved when 
$n_{b:(2.2)}^{(\mathcal{D})} = n_{b:(2.4)}^{(\mathcal{D})} = n_{b:(4.0)}^{(\mathcal{D})} = n_{b:(4.2)}^{(\mathcal{D})} = n_{b:(4.4)}^{(\mathcal{D})} =0$,
and the corresponding minimum energy is
\begin{align}
E=(2-8J)N_p. 
\label{eq:idea}
\end{align}
In this case, the equalities in Eqs.~(\ref{eq:gauss_cluster}), (\ref{eq:constraint2}) and (\ref{eq:constraint3})
hold, i.e.,
\begin{align}
2N_2^{(\mathcal{D})} &+ 4N_4^{(\mathcal{D})} = n_b^{(\mathcal{D})},\label{eq:final_const1}\\
3n_{i:(2.4)}^{(\mathcal{D})}&=n_i^{(\mathcal{D})(2.0)}+n_{i:(2.2)}^{(\mathcal{D})}+n_{i:(2.4)}^{(\mathcal{D})}, \label{eq:final_const2}\\ 
4N^{\mathrm{vac}}&=\sum_{\mathcal{D}}n_{b:(2.0)}^{(\mathcal{D})}. \label{eq:final_const3}
\end{align}
In fact, the ordered phase in Fig.~\ref{fig:apeC} satisfied Eqs.~(\ref{eq:idea})-(\ref{eq:final_const3}). 
Therefore it is the ground state spin configurations for $J>1/4$.

\section{Transfer matrix method}
\begin{figure}[]
\begin{center}
\includegraphics[width=\hsize]{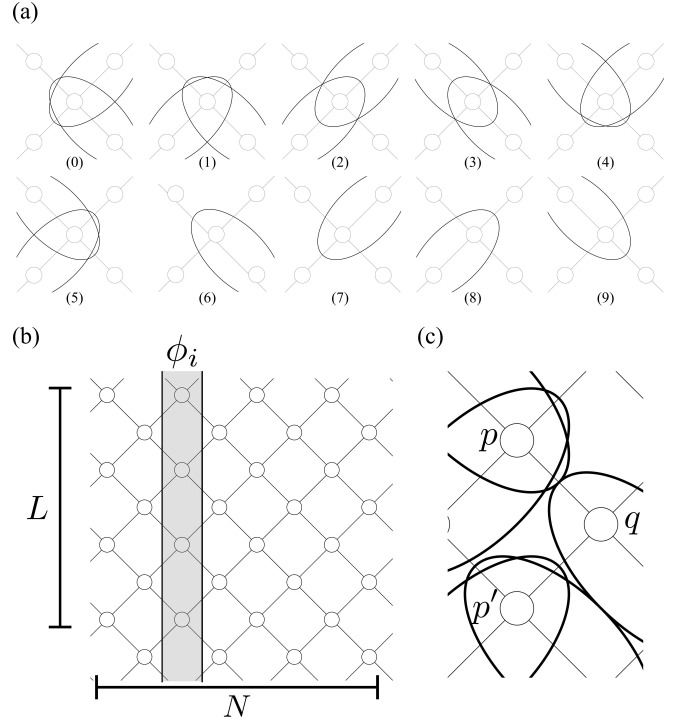}
\end{center}
\caption{(a) All possible ten types of plaquettes in tCSL. (b) Schematic picture of the lattice considered in this Appendix. 
(c)An example configuration and definition of $p, p'$ and $q$. The configuration corresponds to $p=5$ $p'=1$ and $q=6$.}
\label{fig:transfer_matrix}
\end{figure}

In this appendix, we summarize the transfer matrix method we used for the estimation of residual entropy of tCSL. 
We consider the lattice geometry shown in Fig.~\ref{fig:transfer_matrix}(b).
The lattice is composed of $LN$ plaquettes with $L$ rows and $N$ columns, and we impose periodic boundary conditions on both directions.
In this method, we obtain the total number of trimer patterns by identifying the possible trimer configurations column by column.

Specifically, one plaquette can be covered by a trimer in ten different ways, as shown in Fig.~\ref{fig:transfer_matrix}(a).
Let us make a numbering of all possible trimer configurations on the $i-$th column, and denote it as $\phi_i$.
Since there are ten possible configurations for each plaquette, and each column consists of $L$ plaquettes, $\phi_i$ takes $0, 1, \cdots 10^{L}-1$.
However, the trimer configuration on $i$-th column restricts the possible configuration on the $(i+1)$-th column.
As shown in Fig.~\ref{fig:transfer_matrix}(c), type $q$ trimer on a plaquette on the $i$-th column restricts the possible trimer types on adjacent
plaquettes on the $(i+1)$-th column. Possible combinations of trimer types on a triplet of plaquettes, ($p,p',q$) is summarized in a table~\ref{table:possible}.
\begin{table}[htb]
\begin{tabular}{ccc}
p & p' & q \\ \hline \hline
(0,1,3) & (1,3,5) & 9 \\
& (6,8,9) & 9 \\ \hline
(2,4,5) & (0,2,4) & 8 \\
& (1,3,5) & (6,7) \\
& (6,8,9) & (6,7) \\ \hline
6 & 7 & 5 \\
& (6,8,9) & (3,4) \\ \hline
(7,8,9) & (6,8,9) & (0,6,7) \\
& 7 & (1,2) \\ 
& (0,2,4) & 8 \\
& (1,3,5) & (6,7) \\ \hline
\end{tabular}
\caption{Each number represents type of plaquettes in Fig.~\ref{fig:transfer_matrix}(a). Combinations of a row are possible configuration for $p, p'$ and $q$. }
\label{table:possible}
\end{table}

We account for this restriction by introducing the transfer matrix $A^{(L)}$, whose element, $A_{\phi_i,\phi_{i+1} }^{(L)}$, takes 1, if the configurations $\phi_i$ and $\phi_{i+1}$ are compatible, and 0, otherwise. In terms of this transfer matrix, the total number of trimer configurations, $W^{(L,N)}$, can be obtained as
\begin{align}
W^{(L,N)}=&\sum_{\phi_1}\cdots\sum_{\phi_N}A_{\phi_1,{\phi_2} }^{(L)}\cdots A_{\phi_N,{\phi_1} }^{(L)}\nonumber\\
&=\rm{tr}[A_{\phi_i,\phi_{i+1} }^{(L)}]^N=\left(\lambda_L\right)^N,
\end{align}
for sufficiently large $N$. Here, $\lambda_L$ is the largest eigenvalue of $A^{(L)}$. By taking $N\to\infty$, the residual entropy per site is evaluated as 
\begin{align}
S_0= \lim_{N\rightarrow \infty}\frac{1}{LN}\ln(W^{(L,N)}) = \frac{1}{L}\ln\lambda_L.
\end{align}
We numerically evaluate $\lambda_L$ up to $L=7$ and obtain the residual entropy as listed in table~\ref{table:residualentropy}.

\bibliographystyle{apsrev4-1}
\bibliography{checkerboard}

\begin{thebibliography}{54}%
\makeatletter
\providecommand \@ifxundefined [1]{%
 \@ifx{#1\undefined}
}%
\providecommand \@ifnum [1]{%
 \ifnum #1\expandafter \@firstoftwo
 \else \expandafter \@secondoftwo
 \fi
}%
\providecommand \@ifx [1]{%
 \ifx #1\expandafter \@firstoftwo
 \else \expandafter \@secondoftwo
 \fi
}%
\providecommand \natexlab [1]{#1}%
\providecommand \enquote  [1]{``#1''}%
\providecommand \bibnamefont  [1]{#1}%
\providecommand \bibfnamefont [1]{#1}%
\providecommand \citenamefont [1]{#1}%
\providecommand \href@noop [0]{\@secondoftwo}%
\providecommand \href [0]{\begingroup \@sanitize@url \@href}%
\providecommand \@href[1]{\@@startlink{#1}\@@href}%
\providecommand \@@href[1]{\endgroup#1\@@endlink}%
\providecommand \@sanitize@url [0]{\catcode `\\12\catcode `\$12\catcode
  `\&12\catcode `\#12\catcode `\^12\catcode `\_12\catcode `\%12\relax}%
\providecommand \@@startlink[1]{}%
\providecommand \@@endlink[0]{}%
\providecommand \url  [0]{\begingroup\@sanitize@url \@url }%
\providecommand \@url [1]{\endgroup\@href {#1}{\urlprefix }}%
\providecommand \urlprefix  [0]{URL }%
\providecommand \Eprint [0]{\href }%
\providecommand \doibase [0]{http://dx.doi.org/}%
\providecommand \selectlanguage [0]{\@gobble}%
\providecommand \bibinfo  [0]{\@secondoftwo}%
\providecommand \bibfield  [0]{\@secondoftwo}%
\providecommand \translation [1]{[#1]}%
\providecommand \BibitemOpen [0]{}%
\providecommand \bibitemStop [0]{}%
\providecommand \bibitemNoStop [0]{.\EOS\space}%
\providecommand \EOS [0]{\spacefactor3000\relax}%
\providecommand \BibitemShut  [1]{\csname bibitem#1\endcsname}%
\let\auto@bib@innerbib\@empty
\bibitem [{\citenamefont {Bethe}(1931)}]{Bethe1931}%
  \BibitemOpen
  \bibfield  {author} {\bibinfo {author} {\bibfnamefont {H.}~\bibnamefont
  {Bethe}},\ }\href {\doibase 10.1007/BF01341708} {\bibfield  {journal}
  {\bibinfo  {journal} {Zeitschrift f{\"u}r Physik}\ }\textbf {\bibinfo
  {volume} {71}},\ \bibinfo {pages} {205} (\bibinfo {year} {1931})}\BibitemShut
  {NoStop}%
\bibitem [{\citenamefont {Mourigal}\ \emph {et~al.}(2013)\citenamefont
  {Mourigal}, \citenamefont {Enderle}, \citenamefont {Kl{\"o}pperpieper},
  \citenamefont {Caux}, \citenamefont {Stunault},\ and\ \citenamefont
  {R{\o}nnow}}]{mourigal2013fractional}%
  \BibitemOpen
  \bibfield  {author} {\bibinfo {author} {\bibfnamefont {M.}~\bibnamefont
  {Mourigal}}, \bibinfo {author} {\bibfnamefont {M.}~\bibnamefont {Enderle}},
  \bibinfo {author} {\bibfnamefont {A.}~\bibnamefont {Kl{\"o}pperpieper}},
  \bibinfo {author} {\bibfnamefont {J.-S.}\ \bibnamefont {Caux}}, \bibinfo
  {author} {\bibfnamefont {A.}~\bibnamefont {Stunault}}, \ and\ \bibinfo
  {author} {\bibfnamefont {H.~M.}\ \bibnamefont {R{\o}nnow}},\ }\href
  {https://doi.org/10.1038/nphys2652} {\bibfield  {journal} {\bibinfo
  {journal} {Nature Physics}\ }\textbf {\bibinfo {volume} {9}},\ \bibinfo
  {pages} {435} (\bibinfo {year} {2013})}\BibitemShut {NoStop}%
\bibitem [{\citenamefont {Ogata}\ and\ \citenamefont
  {Shiba}(1990)}]{PhysRevB.41.2326}%
  \BibitemOpen
  \bibfield  {author} {\bibinfo {author} {\bibfnamefont {M.}~\bibnamefont
  {Ogata}}\ and\ \bibinfo {author} {\bibfnamefont {H.}~\bibnamefont {Shiba}},\
  }\href {\doibase 10.1103/PhysRevB.41.2326} {\bibfield  {journal} {\bibinfo
  {journal} {Phys. Rev. B}\ }\textbf {\bibinfo {volume} {41}},\ \bibinfo
  {pages} {2326} (\bibinfo {year} {1990})}\BibitemShut {NoStop}%
\bibitem [{\citenamefont {Kitaev}(2001)}]{kitaev2001unpaired}%
  \BibitemOpen
  \bibfield  {author} {\bibinfo {author} {\bibfnamefont {A.~Y.}\ \bibnamefont
  {Kitaev}},\ }\href {https://doi.org/10.1070%2F1063-7869%2F44%2F10s%2Fs29}
  {\bibfield  {journal} {\bibinfo  {journal} {Physics-Uspekhi}\ }\textbf
  {\bibinfo {volume} {44}},\ \bibinfo {pages} {131} (\bibinfo {year}
  {2001})}\BibitemShut {NoStop}%
\bibitem [{\citenamefont {Laughlin}(1983)}]{laughlin1983anomalous}%
  \BibitemOpen
  \bibfield  {author} {\bibinfo {author} {\bibfnamefont {R.~B.}\ \bibnamefont
  {Laughlin}},\ }\href {https://link.aps.org/doi/10.1103/PhysRevLett.50.1395}
  {\bibfield  {journal} {\bibinfo  {journal} {Physical Review Letters}\
  }\textbf {\bibinfo {volume} {50}},\ \bibinfo {pages} {1395} (\bibinfo {year}
  {1983})}\BibitemShut {NoStop}%
\bibitem [{\citenamefont {Tsui}\ \emph {et~al.}(1982)\citenamefont {Tsui},
  \citenamefont {Stormer},\ and\ \citenamefont {Gossard}}]{tsui1982two}%
  \BibitemOpen
  \bibfield  {author} {\bibinfo {author} {\bibfnamefont {D.~C.}\ \bibnamefont
  {Tsui}}, \bibinfo {author} {\bibfnamefont {H.~L.}\ \bibnamefont {Stormer}}, \
  and\ \bibinfo {author} {\bibfnamefont {A.~C.}\ \bibnamefont {Gossard}},\
  }\href {https://link.aps.org/doi/10.1103/PhysRevLett.48.1559} {\bibfield
  {journal} {\bibinfo  {journal} {Physical Review Letters}\ }\textbf {\bibinfo
  {volume} {48}},\ \bibinfo {pages} {1559} (\bibinfo {year}
  {1982})}\BibitemShut {NoStop}%
\bibitem [{\citenamefont {Kitaev}(2006)}]{kitaev2006anyons}%
  \BibitemOpen
  \bibfield  {author} {\bibinfo {author} {\bibfnamefont {A.}~\bibnamefont
  {Kitaev}},\ }\href {http://resolver.caltech.edu/CaltechAUTHORS:KITaop06}
  {\bibfield  {journal} {\bibinfo  {journal} {Annals of Physics}\ }\textbf
  {\bibinfo {volume} {321}},\ \bibinfo {pages} {2} (\bibinfo {year}
  {2006})}\BibitemShut {NoStop}%
\bibitem [{\citenamefont {Knolle}\ \emph {et~al.}(2014)\citenamefont {Knolle},
  \citenamefont {Kovrizhin}, \citenamefont {Chalker},\ and\ \citenamefont
  {Moessner}}]{knolle2014dynamics}%
  \BibitemOpen
  \bibfield  {author} {\bibinfo {author} {\bibfnamefont {J.}~\bibnamefont
  {Knolle}}, \bibinfo {author} {\bibfnamefont {D.~L.}\ \bibnamefont
  {Kovrizhin}}, \bibinfo {author} {\bibfnamefont {J.~T.}\ \bibnamefont
  {Chalker}}, \ and\ \bibinfo {author} {\bibfnamefont {R.}~\bibnamefont
  {Moessner}},\ }\href {\doibase 10.1103/PhysRevLett.112.207203} {\bibfield
  {journal} {\bibinfo  {journal} {Phys. Rev. Lett.}\ }\textbf {\bibinfo
  {volume} {112}},\ \bibinfo {pages} {207203} (\bibinfo {year}
  {2014})}\BibitemShut {NoStop}%
\bibitem [{\citenamefont {Castelnovo}\ \emph {et~al.}(2012)\citenamefont
  {Castelnovo}, \citenamefont {Moessner},\ and\ \citenamefont
  {Sondhi}}]{doi:10.1146/annurev-conmatphys-020911-125058}%
  \BibitemOpen
  \bibfield  {author} {\bibinfo {author} {\bibfnamefont {C.}~\bibnamefont
  {Castelnovo}}, \bibinfo {author} {\bibfnamefont {R.}~\bibnamefont
  {Moessner}}, \ and\ \bibinfo {author} {\bibfnamefont {S.}~\bibnamefont
  {Sondhi}},\ }\href {\doibase 10.1146/annurev-conmatphys-020911-125058}
  {\bibfield  {journal} {\bibinfo  {journal} {Annual Review of Condensed Matter
  Physics}\ }\textbf {\bibinfo {volume} {3}},\ \bibinfo {pages} {35} (\bibinfo
  {year} {2012})}\BibitemShut {NoStop}%
\bibitem [{\citenamefont {Balents}(2010)}]{balents2010balents}%
  \BibitemOpen
  \bibfield  {author} {\bibinfo {author} {\bibfnamefont {L.}~\bibnamefont
  {Balents}},\ }\href {https://doi.org/10.1038/nature08917} {\bibfield
  {journal} {\bibinfo  {journal} {Nature (London)}\ }\textbf {\bibinfo {volume}
  {464}},\ \bibinfo {pages} {199} (\bibinfo {year} {2010})}\BibitemShut
  {NoStop}%
\bibitem [{\citenamefont {Tokiwa}\ \emph {et~al.}(2016)\citenamefont {Tokiwa},
  \citenamefont {Yamashita}, \citenamefont {Udagawa}, \citenamefont {Kittaka},
  \citenamefont {Sakakibara}, \citenamefont {Terazawa}, \citenamefont
  {Shimoyama}, \citenamefont {Terashima}, \citenamefont {Yasui}, \citenamefont
  {Shibauchi},\ and\ \citenamefont {Matsuda}}]{tokiwa2016tokiwa}%
  \BibitemOpen
  \bibfield  {author} {\bibinfo {author} {\bibfnamefont {Y.}~\bibnamefont
  {Tokiwa}}, \bibinfo {author} {\bibfnamefont {T.}~\bibnamefont {Yamashita}},
  \bibinfo {author} {\bibfnamefont {M.}~\bibnamefont {Udagawa}}, \bibinfo
  {author} {\bibfnamefont {S.}~\bibnamefont {Kittaka}}, \bibinfo {author}
  {\bibfnamefont {T.}~\bibnamefont {Sakakibara}}, \bibinfo {author}
  {\bibfnamefont {D.}~\bibnamefont {Terazawa}}, \bibinfo {author}
  {\bibfnamefont {Y.}~\bibnamefont {Shimoyama}}, \bibinfo {author}
  {\bibfnamefont {T.}~\bibnamefont {Terashima}}, \bibinfo {author}
  {\bibfnamefont {Y.}~\bibnamefont {Yasui}}, \bibinfo {author} {\bibfnamefont
  {T.}~\bibnamefont {Shibauchi}}, \ and\ \bibinfo {author} {\bibfnamefont
  {Y.}~\bibnamefont {Matsuda}},\ }\href {https://doi.org/10.1038/ncomms10807}
  {\bibfield  {journal} {\bibinfo  {journal} {Nat. Commun.}\ }\textbf {\bibinfo
  {volume} {7}},\ \bibinfo {pages} {10807} (\bibinfo {year}
  {2016})}\BibitemShut {NoStop}%
\bibitem [{\citenamefont {Pan}\ \emph {et~al.}(2016)\citenamefont {Pan},
  \citenamefont {Laurita}, \citenamefont {Ross}, \citenamefont {Gaulin},\ and\
  \citenamefont {Armitage}}]{pan2016measure}%
  \BibitemOpen
  \bibfield  {author} {\bibinfo {author} {\bibfnamefont {L.}~\bibnamefont
  {Pan}}, \bibinfo {author} {\bibfnamefont {N.}~\bibnamefont {Laurita}},
  \bibinfo {author} {\bibfnamefont {K.~A.}\ \bibnamefont {Ross}}, \bibinfo
  {author} {\bibfnamefont {B.~D.}\ \bibnamefont {Gaulin}}, \ and\ \bibinfo
  {author} {\bibfnamefont {N.}~\bibnamefont {Armitage}},\ }\href
  {https://doi.org/10.1038/nphys3608} {\bibfield  {journal} {\bibinfo
  {journal} {Nature Physics}\ }\textbf {\bibinfo {volume} {12}},\ \bibinfo
  {pages} {361} (\bibinfo {year} {2016})}\BibitemShut {NoStop}%
\bibitem [{\citenamefont {Ross}\ \emph {et~al.}(2011)\citenamefont {Ross},
  \citenamefont {Savary}, \citenamefont {Gaulin},\ and\ \citenamefont
  {Balents}}]{PhysRevX.1.021002}%
  \BibitemOpen
  \bibfield  {author} {\bibinfo {author} {\bibfnamefont {K.~A.}\ \bibnamefont
  {Ross}}, \bibinfo {author} {\bibfnamefont {L.}~\bibnamefont {Savary}},
  \bibinfo {author} {\bibfnamefont {B.~D.}\ \bibnamefont {Gaulin}}, \ and\
  \bibinfo {author} {\bibfnamefont {L.}~\bibnamefont {Balents}},\ }\href
  {\doibase 10.1103/PhysRevX.1.021002} {\bibfield  {journal} {\bibinfo
  {journal} {Phys. Rev. X}\ }\textbf {\bibinfo {volume} {1}},\ \bibinfo {pages}
  {021002} (\bibinfo {year} {2011})}\BibitemShut {NoStop}%
\bibitem [{\citenamefont {Kimura}\ \emph {et~al.}(2013)\citenamefont {Kimura},
  \citenamefont {Nakatsuji}, \citenamefont {Wen}, \citenamefont {Broholm},
  \citenamefont {Stone}, \citenamefont {Nishibori},\ and\ \citenamefont
  {Sawa}}]{kimura2013quantum}%
  \BibitemOpen
  \bibfield  {author} {\bibinfo {author} {\bibfnamefont {K.}~\bibnamefont
  {Kimura}}, \bibinfo {author} {\bibfnamefont {S.}~\bibnamefont {Nakatsuji}},
  \bibinfo {author} {\bibfnamefont {J.}~\bibnamefont {Wen}}, \bibinfo {author}
  {\bibfnamefont {C.}~\bibnamefont {Broholm}}, \bibinfo {author} {\bibfnamefont
  {M.}~\bibnamefont {Stone}}, \bibinfo {author} {\bibfnamefont
  {E.}~\bibnamefont {Nishibori}}, \ and\ \bibinfo {author} {\bibfnamefont
  {H.}~\bibnamefont {Sawa}},\ }\href {https://doi.org/10.1038/ncomms2914}
  {\bibfield  {journal} {\bibinfo  {journal} {Nature Communications}\ }\textbf
  {\bibinfo {volume} {4}},\ \bibinfo {pages} {1934} (\bibinfo {year}
  {2013})}\BibitemShut {NoStop}%
\bibitem [{\citenamefont {Sibille}\ \emph {et~al.}(2018)\citenamefont
  {Sibille}, \citenamefont {Gauthier}, \citenamefont {Yan}, \citenamefont
  {Ciomaga~Hatnean}, \citenamefont {Ollivier}, \citenamefont {Winn},
  \citenamefont {Filges}, \citenamefont {Balakrishnan}, \citenamefont
  {Kenzelmann}, \citenamefont {Shannon},\ and\ \citenamefont
  {Fennell}}]{Sibille2018}%
  \BibitemOpen
  \bibfield  {author} {\bibinfo {author} {\bibfnamefont {R.}~\bibnamefont
  {Sibille}}, \bibinfo {author} {\bibfnamefont {N.}~\bibnamefont {Gauthier}},
  \bibinfo {author} {\bibfnamefont {H.}~\bibnamefont {Yan}}, \bibinfo {author}
  {\bibfnamefont {M.}~\bibnamefont {Ciomaga~Hatnean}}, \bibinfo {author}
  {\bibfnamefont {J.}~\bibnamefont {Ollivier}}, \bibinfo {author}
  {\bibfnamefont {B.}~\bibnamefont {Winn}}, \bibinfo {author} {\bibfnamefont
  {U.}~\bibnamefont {Filges}}, \bibinfo {author} {\bibfnamefont
  {G.}~\bibnamefont {Balakrishnan}}, \bibinfo {author} {\bibfnamefont
  {M.}~\bibnamefont {Kenzelmann}}, \bibinfo {author} {\bibfnamefont
  {N.}~\bibnamefont {Shannon}}, \ and\ \bibinfo {author} {\bibfnamefont
  {T.}~\bibnamefont {Fennell}},\ }\href {\doibase 10.1038/s41567-018-0116-x}
  {\bibfield  {journal} {\bibinfo  {journal} {Nature Physics}\ }\textbf
  {\bibinfo {volume} {14}},\ \bibinfo {pages} {711} (\bibinfo {year}
  {2018})}\BibitemShut {NoStop}%
\bibitem [{\citenamefont {Udagawa}\ and\ \citenamefont
  {Moessner}(2019)}]{PhysRevLett.122.117201}%
  \BibitemOpen
  \bibfield  {author} {\bibinfo {author} {\bibfnamefont {M.}~\bibnamefont
  {Udagawa}}\ and\ \bibinfo {author} {\bibfnamefont {R.}~\bibnamefont
  {Moessner}},\ }\href {\doibase 10.1103/PhysRevLett.122.117201} {\bibfield
  {journal} {\bibinfo  {journal} {Phys. Rev. Lett.}\ }\textbf {\bibinfo
  {volume} {122}},\ \bibinfo {pages} {117201} (\bibinfo {year}
  {2019})}\BibitemShut {NoStop}%
\bibitem [{\citenamefont {Wan}\ \emph {et~al.}(2016)\citenamefont {Wan},
  \citenamefont {Carrasquilla},\ and\ \citenamefont {Melko}}]{wan2016spinon}%
  \BibitemOpen
  \bibfield  {author} {\bibinfo {author} {\bibfnamefont {Y.}~\bibnamefont
  {Wan}}, \bibinfo {author} {\bibfnamefont {J.}~\bibnamefont {Carrasquilla}}, \
  and\ \bibinfo {author} {\bibfnamefont {R.~G.}\ \bibnamefont {Melko}},\ }\href
  {https://link.aps.org/doi/10.1103/PhysRevLett.116.167202} {\bibfield
  {journal} {\bibinfo  {journal} {Physical review letters}\ }\textbf {\bibinfo
  {volume} {116}},\ \bibinfo {pages} {167202} (\bibinfo {year}
  {2016})}\BibitemShut {NoStop}%
\bibitem [{\citenamefont {Kourtis}\ and\ \citenamefont
  {Castelnovo}(2016)}]{kourtis2016free}%
  \BibitemOpen
  \bibfield  {author} {\bibinfo {author} {\bibfnamefont {S.}~\bibnamefont
  {Kourtis}}\ and\ \bibinfo {author} {\bibfnamefont {C.}~\bibnamefont
  {Castelnovo}},\ }\href {\doibase 10.1103/PhysRevB.94.104401} {\bibfield
  {journal} {\bibinfo  {journal} {Phys. Rev. B}\ }\textbf {\bibinfo {volume}
  {94}},\ \bibinfo {pages} {104401} (\bibinfo {year} {2016})}\BibitemShut
  {NoStop}%
\bibitem [{\citenamefont {Huang}\ \emph {et~al.}(2018)\citenamefont {Huang},
  \citenamefont {Deng}, \citenamefont {Wan},\ and\ \citenamefont
  {Meng}}]{PhysRevLett.120.167202}%
  \BibitemOpen
  \bibfield  {author} {\bibinfo {author} {\bibfnamefont {C.-J.}\ \bibnamefont
  {Huang}}, \bibinfo {author} {\bibfnamefont {Y.}~\bibnamefont {Deng}},
  \bibinfo {author} {\bibfnamefont {Y.}~\bibnamefont {Wan}}, \ and\ \bibinfo
  {author} {\bibfnamefont {Z.~Y.}\ \bibnamefont {Meng}},\ }\href {\doibase
  10.1103/PhysRevLett.120.167202} {\bibfield  {journal} {\bibinfo  {journal}
  {Phys. Rev. Lett.}\ }\textbf {\bibinfo {volume} {120}},\ \bibinfo {pages}
  {167202} (\bibinfo {year} {2018})}\BibitemShut {NoStop}%
\bibitem [{\citenamefont {Chen}(2017)}]{PhysRevB.96.195127}%
  \BibitemOpen
  \bibfield  {author} {\bibinfo {author} {\bibfnamefont {G.}~\bibnamefont
  {Chen}},\ }\href {\doibase 10.1103/PhysRevB.96.195127} {\bibfield  {journal}
  {\bibinfo  {journal} {Phys. Rev. B}\ }\textbf {\bibinfo {volume} {96}},\
  \bibinfo {pages} {195127} (\bibinfo {year} {2017})}\BibitemShut {NoStop}%
\bibitem [{\citenamefont {Banerjee}\ \emph {et~al.}(2016)\citenamefont
  {Banerjee}, \citenamefont {Bridges}, \citenamefont {Yan}, \citenamefont
  {Aczel}, \citenamefont {Li}, \citenamefont {Stone}, \citenamefont {Granroth},
  \citenamefont {Lumsden}, \citenamefont {Yiu}, \citenamefont {Knolle},
  \citenamefont {Bhattacharjee}, \citenamefont {Kovrizhin}, \citenamefont
  {Moessner}, \citenamefont {Tennant}, \citenamefont {Mandrus},\ and\
  \citenamefont {Nagler}}]{banerjee2016proximate}%
  \BibitemOpen
  \bibfield  {author} {\bibinfo {author} {\bibfnamefont {A.}~\bibnamefont
  {Banerjee}}, \bibinfo {author} {\bibfnamefont {C.~A.}\ \bibnamefont
  {Bridges}}, \bibinfo {author} {\bibfnamefont {J.-Q.}\ \bibnamefont {Yan}},
  \bibinfo {author} {\bibfnamefont {A.~A.}\ \bibnamefont {Aczel}}, \bibinfo
  {author} {\bibfnamefont {L.}~\bibnamefont {Li}}, \bibinfo {author}
  {\bibfnamefont {M.~B.}\ \bibnamefont {Stone}}, \bibinfo {author}
  {\bibfnamefont {G.~E.}\ \bibnamefont {Granroth}}, \bibinfo {author}
  {\bibfnamefont {M.~D.}\ \bibnamefont {Lumsden}}, \bibinfo {author}
  {\bibfnamefont {Y.}~\bibnamefont {Yiu}}, \bibinfo {author} {\bibfnamefont
  {J.}~\bibnamefont {Knolle}}, \bibinfo {author} {\bibfnamefont
  {S.}~\bibnamefont {Bhattacharjee}}, \bibinfo {author} {\bibfnamefont {D.~L.}\
  \bibnamefont {Kovrizhin}}, \bibinfo {author} {\bibfnamefont {R.}~\bibnamefont
  {Moessner}}, \bibinfo {author} {\bibfnamefont {D.~A.}\ \bibnamefont
  {Tennant}}, \bibinfo {author} {\bibfnamefont {D.~G.}\ \bibnamefont
  {Mandrus}}, \ and\ \bibinfo {author} {\bibfnamefont {S.~E.}\ \bibnamefont
  {Nagler}},\ }\href {https://doi.org/10.1038/nmat4604} {\bibfield  {journal}
  {\bibinfo  {journal} {Nature Materials}\ }\textbf {\bibinfo {volume} {15}},\
  \bibinfo {pages} {733} (\bibinfo {year} {2016})}\BibitemShut {NoStop}%
\bibitem [{\citenamefont {Banerjee}\ \emph {et~al.}(2017)\citenamefont
  {Banerjee}, \citenamefont {Yan}, \citenamefont {Knolle}, \citenamefont
  {Bridges}, \citenamefont {Stone}, \citenamefont {Lumsden}, \citenamefont
  {Mandrus}, \citenamefont {Tennant}, \citenamefont {Moessner},\ and\
  \citenamefont {Nagler}}]{banerjee2017neutron}%
  \BibitemOpen
  \bibfield  {author} {\bibinfo {author} {\bibfnamefont {A.}~\bibnamefont
  {Banerjee}}, \bibinfo {author} {\bibfnamefont {J.}~\bibnamefont {Yan}},
  \bibinfo {author} {\bibfnamefont {J.}~\bibnamefont {Knolle}}, \bibinfo
  {author} {\bibfnamefont {C.~A.}\ \bibnamefont {Bridges}}, \bibinfo {author}
  {\bibfnamefont {M.~B.}\ \bibnamefont {Stone}}, \bibinfo {author}
  {\bibfnamefont {M.~D.}\ \bibnamefont {Lumsden}}, \bibinfo {author}
  {\bibfnamefont {D.~G.}\ \bibnamefont {Mandrus}}, \bibinfo {author}
  {\bibfnamefont {D.~A.}\ \bibnamefont {Tennant}}, \bibinfo {author}
  {\bibfnamefont {R.}~\bibnamefont {Moessner}}, \ and\ \bibinfo {author}
  {\bibfnamefont {S.~E.}\ \bibnamefont {Nagler}},\ }\href
  {https://science.sciencemag.org/content/356/6342/1055} {\bibfield  {journal}
  {\bibinfo  {journal} {Science}\ }\textbf {\bibinfo {volume} {356}},\ \bibinfo
  {pages} {1055} (\bibinfo {year} {2017})}\BibitemShut {NoStop}%
\bibitem [{\citenamefont {{Do}}\ \emph {et~al.}(2017)\citenamefont {{Do}},
  \citenamefont {{Park}}, \citenamefont {{Yoshitake}}, \citenamefont {{Nasu}},
  \citenamefont {{Motome}}, \citenamefont {{Kwon}}, \citenamefont {{Adroja}},
  \citenamefont {{Voneshen}}, \citenamefont {{Kim}}, \citenamefont {{Jang}},
  \citenamefont {{Park}}, \citenamefont {{Choi}},\ and\ \citenamefont
  {{Ji}}}]{2017NatPh..13.1079D}%
  \BibitemOpen
  \bibfield  {author} {\bibinfo {author} {\bibfnamefont {S.-H.}\ \bibnamefont
  {{Do}}}, \bibinfo {author} {\bibfnamefont {S.-Y.}\ \bibnamefont {{Park}}},
  \bibinfo {author} {\bibfnamefont {J.}~\bibnamefont {{Yoshitake}}}, \bibinfo
  {author} {\bibfnamefont {J.}~\bibnamefont {{Nasu}}}, \bibinfo {author}
  {\bibfnamefont {Y.}~\bibnamefont {{Motome}}}, \bibinfo {author}
  {\bibfnamefont {Y.}~\bibnamefont {{Kwon}}}, \bibinfo {author} {\bibfnamefont
  {D.}~\bibnamefont {{Adroja}}}, \bibinfo {author} {\bibfnamefont
  {D.}~\bibnamefont {{Voneshen}}}, \bibinfo {author} {\bibfnamefont
  {K.}~\bibnamefont {{Kim}}}, \bibinfo {author} {\bibfnamefont {T.-H.}\
  \bibnamefont {{Jang}}}, \bibinfo {author} {\bibfnamefont {J.-H.}\
  \bibnamefont {{Park}}}, \bibinfo {author} {\bibfnamefont {K.-Y.}\
  \bibnamefont {{Choi}}}, \ and\ \bibinfo {author} {\bibfnamefont
  {S.}~\bibnamefont {{Ji}}},\ }\href {\doibase 10.1038/nphys4264} {\bibfield
  {journal} {\bibinfo  {journal} {Nature Physics}\ }\textbf {\bibinfo {volume}
  {13}},\ \bibinfo {pages} {1079} (\bibinfo {year} {2017})}\BibitemShut
  {NoStop}%
\bibitem [{\citenamefont {Kasahara}\ \emph {et~al.}(2018)\citenamefont
  {Kasahara}, \citenamefont {Ohnishi}, \citenamefont {Mizukami}, \citenamefont
  {Tanaka}, \citenamefont {Ma}, \citenamefont {Sugii}, \citenamefont {Kurita},
  \citenamefont {Tanaka}, \citenamefont {Nasu}, \citenamefont {Motome},
  \citenamefont {Shibauchi},\ and\ \citenamefont
  {Matsuda}}]{kasahara2018majorana}%
  \BibitemOpen
  \bibfield  {author} {\bibinfo {author} {\bibfnamefont {Y.}~\bibnamefont
  {Kasahara}}, \bibinfo {author} {\bibfnamefont {T.}~\bibnamefont {Ohnishi}},
  \bibinfo {author} {\bibfnamefont {Y.}~\bibnamefont {Mizukami}}, \bibinfo
  {author} {\bibfnamefont {O.}~\bibnamefont {Tanaka}}, \bibinfo {author}
  {\bibfnamefont {S.}~\bibnamefont {Ma}}, \bibinfo {author} {\bibfnamefont
  {K.}~\bibnamefont {Sugii}}, \bibinfo {author} {\bibfnamefont
  {N.}~\bibnamefont {Kurita}}, \bibinfo {author} {\bibfnamefont
  {H.}~\bibnamefont {Tanaka}}, \bibinfo {author} {\bibfnamefont
  {J.}~\bibnamefont {Nasu}}, \bibinfo {author} {\bibfnamefont {Y.}~\bibnamefont
  {Motome}}, \bibinfo {author} {\bibfnamefont {T.}~\bibnamefont {Shibauchi}}, \
  and\ \bibinfo {author} {\bibfnamefont {Y.}~\bibnamefont {Matsuda}},\ }\href
  {https://doi.org/10.1038/s41586-018-0274-0} {\bibfield  {journal} {\bibinfo
  {journal} {Nature}\ }\textbf {\bibinfo {volume} {559}},\ \bibinfo {pages}
  {227} (\bibinfo {year} {2018})}\BibitemShut {NoStop}%
\bibitem [{\citenamefont {Castelnovo}\ \emph {et~al.}(2008)\citenamefont
  {Castelnovo}, \citenamefont {Moessner},\ and\ \citenamefont
  {Sondhi}}]{castelnovo2008magnetic}%
  \BibitemOpen
  \bibfield  {author} {\bibinfo {author} {\bibfnamefont {C.}~\bibnamefont
  {Castelnovo}}, \bibinfo {author} {\bibfnamefont {R.}~\bibnamefont
  {Moessner}}, \ and\ \bibinfo {author} {\bibfnamefont {S.~L.}\ \bibnamefont
  {Sondhi}},\ }\href {https://doi.org/10.1038/nature06433} {\bibfield
  {journal} {\bibinfo  {journal} {Nature}\ }\textbf {\bibinfo {volume} {451}},\
  \bibinfo {pages} {42} (\bibinfo {year} {2008})}\BibitemShut {NoStop}%
\bibitem [{\citenamefont {Sakakibara}\ \emph {et~al.}(2003)\citenamefont
  {Sakakibara}, \citenamefont {Tayama}, \citenamefont {Hiroi}, \citenamefont
  {Matsuhira},\ and\ \citenamefont {Takagi}}]{PhysRevLett.90.207205}%
  \BibitemOpen
  \bibfield  {author} {\bibinfo {author} {\bibfnamefont {T.}~\bibnamefont
  {Sakakibara}}, \bibinfo {author} {\bibfnamefont {T.}~\bibnamefont {Tayama}},
  \bibinfo {author} {\bibfnamefont {Z.}~\bibnamefont {Hiroi}}, \bibinfo
  {author} {\bibfnamefont {K.}~\bibnamefont {Matsuhira}}, \ and\ \bibinfo
  {author} {\bibfnamefont {S.}~\bibnamefont {Takagi}},\ }\href {\doibase
  10.1103/PhysRevLett.90.207205} {\bibfield  {journal} {\bibinfo  {journal}
  {Phys. Rev. Lett.}\ }\textbf {\bibinfo {volume} {90}},\ \bibinfo {pages}
  {207205} (\bibinfo {year} {2003})}\BibitemShut {NoStop}%
\bibitem [{\citenamefont {Mizoguchi}\ \emph {et~al.}(2017)\citenamefont
  {Mizoguchi}, \citenamefont {Jaubert},\ and\ \citenamefont
  {Udagawa}}]{PhysRevLett.119.077207}%
  \BibitemOpen
  \bibfield  {author} {\bibinfo {author} {\bibfnamefont {T.}~\bibnamefont
  {Mizoguchi}}, \bibinfo {author} {\bibfnamefont {L.~D.~C.}\ \bibnamefont
  {Jaubert}}, \ and\ \bibinfo {author} {\bibfnamefont {M.}~\bibnamefont
  {Udagawa}},\ }\href {\doibase 10.1103/PhysRevLett.119.077207} {\bibfield
  {journal} {\bibinfo  {journal} {Phys. Rev. Lett.}\ }\textbf {\bibinfo
  {volume} {119}},\ \bibinfo {pages} {077207} (\bibinfo {year}
  {2017})}\BibitemShut {NoStop}%
\bibitem [{\citenamefont {Udagawa}\ \emph {et~al.}(2016)\citenamefont
  {Udagawa}, \citenamefont {Jaubert}, \citenamefont {Castelnovo},\ and\
  \citenamefont {Moessner}}]{PhysRevB.94.104416}%
  \BibitemOpen
  \bibfield  {author} {\bibinfo {author} {\bibfnamefont {M.}~\bibnamefont
  {Udagawa}}, \bibinfo {author} {\bibfnamefont {L.~D.~C.}\ \bibnamefont
  {Jaubert}}, \bibinfo {author} {\bibfnamefont {C.}~\bibnamefont {Castelnovo}},
  \ and\ \bibinfo {author} {\bibfnamefont {R.}~\bibnamefont {Moessner}},\
  }\href {\doibase 10.1103/PhysRevB.94.104416} {\bibfield  {journal} {\bibinfo
  {journal} {Phys. Rev. B}\ }\textbf {\bibinfo {volume} {94}},\ \bibinfo
  {pages} {104416} (\bibinfo {year} {2016})}\BibitemShut {NoStop}%
\bibitem [{\citenamefont {Rau}\ and\ \citenamefont
  {Gingras}(2016)}]{rau2016spin}%
  \BibitemOpen
  \bibfield  {author} {\bibinfo {author} {\bibfnamefont {J.~G.}\ \bibnamefont
  {Rau}}\ and\ \bibinfo {author} {\bibfnamefont {M.~J.}\ \bibnamefont
  {Gingras}},\ }\href {https://doi.org/10.1038/ncomms12234} {\bibfield
  {journal} {\bibinfo  {journal} {Nature Communications}\ }\textbf {\bibinfo
  {volume} {7}},\ \bibinfo {pages} {12234} (\bibinfo {year}
  {2016})}\BibitemShut {NoStop}%
\bibitem [{\citenamefont {Mizoguchi}\ \emph {et~al.}(2018)\citenamefont
  {Mizoguchi}, \citenamefont {Jaubert}, \citenamefont {Moessner},\ and\
  \citenamefont {Udagawa}}]{mizoguchi2018magnetic}%
  \BibitemOpen
  \bibfield  {author} {\bibinfo {author} {\bibfnamefont {T.}~\bibnamefont
  {Mizoguchi}}, \bibinfo {author} {\bibfnamefont {L.~D.~C.}\ \bibnamefont
  {Jaubert}}, \bibinfo {author} {\bibfnamefont {R.}~\bibnamefont {Moessner}}, \
  and\ \bibinfo {author} {\bibfnamefont {M.}~\bibnamefont {Udagawa}},\ }\href
  {\doibase 10.1103/PhysRevB.98.144446} {\bibfield  {journal} {\bibinfo
  {journal} {Phys. Rev. B}\ }\textbf {\bibinfo {volume} {98}},\ \bibinfo
  {pages} {144446} (\bibinfo {year} {2018})}\BibitemShut {NoStop}%
\bibitem [{\citenamefont {Yan}\ \emph {et~al.}(2018)\citenamefont {Yan},
  \citenamefont {Pohle},\ and\ \citenamefont {Shannon}}]{PhysRevB.98.140402}%
  \BibitemOpen
  \bibfield  {author} {\bibinfo {author} {\bibfnamefont {H.}~\bibnamefont
  {Yan}}, \bibinfo {author} {\bibfnamefont {R.}~\bibnamefont {Pohle}}, \ and\
  \bibinfo {author} {\bibfnamefont {N.}~\bibnamefont {Shannon}},\ }\href
  {\doibase 10.1103/PhysRevB.98.140402} {\bibfield  {journal} {\bibinfo
  {journal} {Phys. Rev. B}\ }\textbf {\bibinfo {volume} {98}},\ \bibinfo
  {pages} {140402(R)} (\bibinfo {year} {2018})}\BibitemShut {NoStop}%
\bibitem [{\citenamefont {Ono}\ \emph {et~al.}(2003)\citenamefont {Ono},
  \citenamefont {Tanaka}, \citenamefont {Aruga~Katori}, \citenamefont
  {Ishikawa}, \citenamefont {Mitamura},\ and\ \citenamefont
  {Goto}}]{PhysRevB.67.104431}%
  \BibitemOpen
  \bibfield  {author} {\bibinfo {author} {\bibfnamefont {T.}~\bibnamefont
  {Ono}}, \bibinfo {author} {\bibfnamefont {H.}~\bibnamefont {Tanaka}},
  \bibinfo {author} {\bibfnamefont {H.}~\bibnamefont {Aruga~Katori}}, \bibinfo
  {author} {\bibfnamefont {F.}~\bibnamefont {Ishikawa}}, \bibinfo {author}
  {\bibfnamefont {H.}~\bibnamefont {Mitamura}}, \ and\ \bibinfo {author}
  {\bibfnamefont {T.}~\bibnamefont {Goto}},\ }\href {\doibase
  10.1103/PhysRevB.67.104431} {\bibfield  {journal} {\bibinfo  {journal} {Phys.
  Rev. B}\ }\textbf {\bibinfo {volume} {67}},\ \bibinfo {pages} {104431}
  (\bibinfo {year} {2003})}\BibitemShut {NoStop}%
\bibitem [{\citenamefont {Miyata}\ \emph {et~al.}(2017)\citenamefont {Miyata},
  \citenamefont {Portugall}, \citenamefont {Nakamura}, \citenamefont
  {Ohgushi},\ and\ \citenamefont {Takeyama}}]{PhysRevB.96.180401}%
  \BibitemOpen
  \bibfield  {author} {\bibinfo {author} {\bibfnamefont {A.}~\bibnamefont
  {Miyata}}, \bibinfo {author} {\bibfnamefont {O.}~\bibnamefont {Portugall}},
  \bibinfo {author} {\bibfnamefont {D.}~\bibnamefont {Nakamura}}, \bibinfo
  {author} {\bibfnamefont {K.}~\bibnamefont {Ohgushi}}, \ and\ \bibinfo
  {author} {\bibfnamefont {S.}~\bibnamefont {Takeyama}},\ }\href {\doibase
  10.1103/PhysRevB.96.180401} {\bibfield  {journal} {\bibinfo  {journal} {Phys.
  Rev. B}\ }\textbf {\bibinfo {volume} {96}},\ \bibinfo {pages} {180401(R)}
  (\bibinfo {year} {2017})}\BibitemShut {NoStop}%
\bibitem [{\citenamefont {Alicea}\ \emph {et~al.}(2009)\citenamefont {Alicea},
  \citenamefont {Chubukov},\ and\ \citenamefont
  {Starykh}}]{PhysRevLett.102.137201}%
  \BibitemOpen
  \bibfield  {author} {\bibinfo {author} {\bibfnamefont {J.}~\bibnamefont
  {Alicea}}, \bibinfo {author} {\bibfnamefont {A.~V.}\ \bibnamefont
  {Chubukov}}, \ and\ \bibinfo {author} {\bibfnamefont {O.~A.}\ \bibnamefont
  {Starykh}},\ }\href {\doibase 10.1103/PhysRevLett.102.137201} {\bibfield
  {journal} {\bibinfo  {journal} {Phys. Rev. Lett.}\ }\textbf {\bibinfo
  {volume} {102}},\ \bibinfo {pages} {137201} (\bibinfo {year}
  {2009})}\BibitemShut {NoStop}%
\bibitem [{\citenamefont {Coletta}\ \emph {et~al.}(2016)\citenamefont
  {Coletta}, \citenamefont {T\'oth}, \citenamefont {Penc},\ and\ \citenamefont
  {Mila}}]{PhysRevB.94.075136}%
  \BibitemOpen
  \bibfield  {author} {\bibinfo {author} {\bibfnamefont {T.}~\bibnamefont
  {Coletta}}, \bibinfo {author} {\bibfnamefont {T.~A.}\ \bibnamefont {T\'oth}},
  \bibinfo {author} {\bibfnamefont {K.}~\bibnamefont {Penc}}, \ and\ \bibinfo
  {author} {\bibfnamefont {F.}~\bibnamefont {Mila}},\ }\href {\doibase
  10.1103/PhysRevB.94.075136} {\bibfield  {journal} {\bibinfo  {journal} {Phys.
  Rev. B}\ }\textbf {\bibinfo {volume} {94}},\ \bibinfo {pages} {075136}
  (\bibinfo {year} {2016})}\BibitemShut {NoStop}%
\bibitem [{\citenamefont {Momoi}\ and\ \citenamefont
  {Totsuka}(2000)}]{PhysRevB.62.15067}%
  \BibitemOpen
  \bibfield  {author} {\bibinfo {author} {\bibfnamefont {T.}~\bibnamefont
  {Momoi}}\ and\ \bibinfo {author} {\bibfnamefont {K.}~\bibnamefont
  {Totsuka}},\ }\href {\doibase 10.1103/PhysRevB.62.15067} {\bibfield
  {journal} {\bibinfo  {journal} {Phys. Rev. B}\ }\textbf {\bibinfo {volume}
  {62}},\ \bibinfo {pages} {15067} (\bibinfo {year} {2000})}\BibitemShut
  {NoStop}%
\bibitem [{\citenamefont {Yamamoto}\ \emph {et~al.}(2014)\citenamefont
  {Yamamoto}, \citenamefont {Marmorini},\ and\ \citenamefont
  {Danshita}}]{PhysRevLett.112.127203}%
  \BibitemOpen
  \bibfield  {author} {\bibinfo {author} {\bibfnamefont {D.}~\bibnamefont
  {Yamamoto}}, \bibinfo {author} {\bibfnamefont {G.}~\bibnamefont {Marmorini}},
  \ and\ \bibinfo {author} {\bibfnamefont {I.}~\bibnamefont {Danshita}},\
  }\href {\doibase 10.1103/PhysRevLett.112.127203} {\bibfield  {journal}
  {\bibinfo  {journal} {Phys. Rev. Lett.}\ }\textbf {\bibinfo {volume} {112}},\
  \bibinfo {pages} {127203} (\bibinfo {year} {2014})}\BibitemShut {NoStop}%
\bibitem [{\citenamefont {Sakai}\ and\ \citenamefont
  {Nakano}(2011)}]{PhysRevB.83.100405}%
  \BibitemOpen
  \bibfield  {author} {\bibinfo {author} {\bibfnamefont {T.}~\bibnamefont
  {Sakai}}\ and\ \bibinfo {author} {\bibfnamefont {H.}~\bibnamefont {Nakano}},\
  }\href {\doibase 10.1103/PhysRevB.83.100405} {\bibfield  {journal} {\bibinfo
  {journal} {Phys. Rev. B}\ }\textbf {\bibinfo {volume} {83}},\ \bibinfo
  {pages} {100405(R)} (\bibinfo {year} {2011})}\BibitemShut {NoStop}%
\bibitem [{\citenamefont {Shirata}\ \emph {et~al.}(2012)\citenamefont
  {Shirata}, \citenamefont {Tanaka}, \citenamefont {Matsuo},\ and\
  \citenamefont {Kindo}}]{PhysRevLett.108.057205}%
  \BibitemOpen
  \bibfield  {author} {\bibinfo {author} {\bibfnamefont {Y.}~\bibnamefont
  {Shirata}}, \bibinfo {author} {\bibfnamefont {H.}~\bibnamefont {Tanaka}},
  \bibinfo {author} {\bibfnamefont {A.}~\bibnamefont {Matsuo}}, \ and\ \bibinfo
  {author} {\bibfnamefont {K.}~\bibnamefont {Kindo}},\ }\href {\doibase
  10.1103/PhysRevLett.108.057205} {\bibfield  {journal} {\bibinfo  {journal}
  {Phys. Rev. Lett.}\ }\textbf {\bibinfo {volume} {108}},\ \bibinfo {pages}
  {057205} (\bibinfo {year} {2012})}\BibitemShut {NoStop}%
\bibitem [{\citenamefont {Susuki}\ \emph {et~al.}(2013)\citenamefont {Susuki},
  \citenamefont {Kurita}, \citenamefont {Tanaka}, \citenamefont {Nojiri},
  \citenamefont {Matsuo}, \citenamefont {Kindo},\ and\ \citenamefont
  {Tanaka}}]{PhysRevLett.110.267201}%
  \BibitemOpen
  \bibfield  {author} {\bibinfo {author} {\bibfnamefont {T.}~\bibnamefont
  {Susuki}}, \bibinfo {author} {\bibfnamefont {N.}~\bibnamefont {Kurita}},
  \bibinfo {author} {\bibfnamefont {T.}~\bibnamefont {Tanaka}}, \bibinfo
  {author} {\bibfnamefont {H.}~\bibnamefont {Nojiri}}, \bibinfo {author}
  {\bibfnamefont {A.}~\bibnamefont {Matsuo}}, \bibinfo {author} {\bibfnamefont
  {K.}~\bibnamefont {Kindo}}, \ and\ \bibinfo {author} {\bibfnamefont
  {H.}~\bibnamefont {Tanaka}},\ }\href {\doibase
  10.1103/PhysRevLett.110.267201} {\bibfield  {journal} {\bibinfo  {journal}
  {Phys. Rev. Lett.}\ }\textbf {\bibinfo {volume} {110}},\ \bibinfo {pages}
  {267201} (\bibinfo {year} {2013})}\BibitemShut {NoStop}%
\bibitem [{\citenamefont {Onizuka}\ \emph {et~al.}(2000)\citenamefont
  {Onizuka}, \citenamefont {Kageyama}, \citenamefont {Narumi}, \citenamefont
  {Kindo}, \citenamefont {Ueda},\ and\ \citenamefont
  {Goto}}]{doi:10.1143/JPSJ.69.1016}%
  \BibitemOpen
  \bibfield  {author} {\bibinfo {author} {\bibfnamefont {K.}~\bibnamefont
  {Onizuka}}, \bibinfo {author} {\bibfnamefont {H.}~\bibnamefont {Kageyama}},
  \bibinfo {author} {\bibfnamefont {Y.}~\bibnamefont {Narumi}}, \bibinfo
  {author} {\bibfnamefont {K.}~\bibnamefont {Kindo}}, \bibinfo {author}
  {\bibfnamefont {Y.}~\bibnamefont {Ueda}}, \ and\ \bibinfo {author}
  {\bibfnamefont {T.}~\bibnamefont {Goto}},\ }\href
  {https://doi.org/10.1143/JPSJ.69.1016} {\bibfield  {journal} {\bibinfo
  {journal} {Journal of the Physical Society of Japan}\ }\textbf {\bibinfo
  {volume} {69}},\ \bibinfo {pages} {1016} (\bibinfo {year}
  {2000})}\BibitemShut {NoStop}%
\bibitem [{\citenamefont {Matsuda}\ \emph {et~al.}(2013)\citenamefont
  {Matsuda}, \citenamefont {Abe}, \citenamefont {Takeyama}, \citenamefont
  {Kageyama}, \citenamefont {Corboz}, \citenamefont {Honecker}, \citenamefont
  {Manmana}, \citenamefont {Foltin}, \citenamefont {Schmidt},\ and\
  \citenamefont {Mila}}]{PhysRevLett.111.137204}%
  \BibitemOpen
  \bibfield  {author} {\bibinfo {author} {\bibfnamefont {Y.~H.}\ \bibnamefont
  {Matsuda}}, \bibinfo {author} {\bibfnamefont {N.}~\bibnamefont {Abe}},
  \bibinfo {author} {\bibfnamefont {S.}~\bibnamefont {Takeyama}}, \bibinfo
  {author} {\bibfnamefont {H.}~\bibnamefont {Kageyama}}, \bibinfo {author}
  {\bibfnamefont {P.}~\bibnamefont {Corboz}}, \bibinfo {author} {\bibfnamefont
  {A.}~\bibnamefont {Honecker}}, \bibinfo {author} {\bibfnamefont {S.~R.}\
  \bibnamefont {Manmana}}, \bibinfo {author} {\bibfnamefont {G.~R.}\
  \bibnamefont {Foltin}}, \bibinfo {author} {\bibfnamefont {K.~P.}\
  \bibnamefont {Schmidt}}, \ and\ \bibinfo {author} {\bibfnamefont
  {F.}~\bibnamefont {Mila}},\ }\href {\doibase 10.1103/PhysRevLett.111.137204}
  {\bibfield  {journal} {\bibinfo  {journal} {Phys. Rev. Lett.}\ }\textbf
  {\bibinfo {volume} {111}},\ \bibinfo {pages} {137204} (\bibinfo {year}
  {2013})}\BibitemShut {NoStop}%
\bibitem [{\citenamefont {Miyata}\ \emph {et~al.}(2013)\citenamefont {Miyata},
  \citenamefont {Takeyama},\ and\ \citenamefont {Ueda}}]{PhysRevB.87.214424}%
  \BibitemOpen
  \bibfield  {author} {\bibinfo {author} {\bibfnamefont {A.}~\bibnamefont
  {Miyata}}, \bibinfo {author} {\bibfnamefont {S.}~\bibnamefont {Takeyama}}, \
  and\ \bibinfo {author} {\bibfnamefont {H.}~\bibnamefont {Ueda}},\ }\href
  {\doibase 10.1103/PhysRevB.87.214424} {\bibfield  {journal} {\bibinfo
  {journal} {Phys. Rev. B}\ }\textbf {\bibinfo {volume} {87}},\ \bibinfo
  {pages} {214424} (\bibinfo {year} {2013})}\BibitemShut {NoStop}%
\bibitem [{\citenamefont {Nakamura}\ \emph {et~al.}(2014)\citenamefont
  {Nakamura}, \citenamefont {Miyata}, \citenamefont {Aida}, \citenamefont
  {Ueda},\ and\ \citenamefont {Takeyama}}]{doi:10.7566/JPSJ.83.113703}%
  \BibitemOpen
  \bibfield  {author} {\bibinfo {author} {\bibfnamefont {D.}~\bibnamefont
  {Nakamura}}, \bibinfo {author} {\bibfnamefont {A.}~\bibnamefont {Miyata}},
  \bibinfo {author} {\bibfnamefont {Y.}~\bibnamefont {Aida}}, \bibinfo {author}
  {\bibfnamefont {H.}~\bibnamefont {Ueda}}, \ and\ \bibinfo {author}
  {\bibfnamefont {S.}~\bibnamefont {Takeyama}},\ }\href
  {https://doi.org/10.7566/JPSJ.83.113703} {\bibfield  {journal} {\bibinfo
  {journal} {Journal of the Physical Society of Japan}\ }\textbf {\bibinfo
  {volume} {83}},\ \bibinfo {pages} {113703} (\bibinfo {year}
  {2014})}\BibitemShut {NoStop}%
\bibitem [{\citenamefont {Nishimoto}\ \emph {et~al.}(2013)\citenamefont
  {Nishimoto}, \citenamefont {Shibata},\ and\ \citenamefont
  {Hotta}}]{nishimoto2013controlling}%
  \BibitemOpen
  \bibfield  {author} {\bibinfo {author} {\bibfnamefont {S.}~\bibnamefont
  {Nishimoto}}, \bibinfo {author} {\bibfnamefont {N.}~\bibnamefont {Shibata}},
  \ and\ \bibinfo {author} {\bibfnamefont {C.}~\bibnamefont {Hotta}},\ }\href
  {https://doi.org/10.1038/ncomms3287} {\bibfield  {journal} {\bibinfo
  {journal} {Nature Communications}\ }\textbf {\bibinfo {volume} {4}},\
  \bibinfo {pages} {2287} (\bibinfo {year} {2013})}\BibitemShut {NoStop}%
\bibitem [{\citenamefont {Lieb}(1967)}]{PhysRevLett.18.692}%
  \BibitemOpen
  \bibfield  {author} {\bibinfo {author} {\bibfnamefont {E.~H.}\ \bibnamefont
  {Lieb}},\ }\href {\doibase 10.1103/PhysRevLett.18.692} {\bibfield  {journal}
  {\bibinfo  {journal} {Phys. Rev. Lett.}\ }\textbf {\bibinfo {volume} {18}},\
  \bibinfo {pages} {692} (\bibinfo {year} {1967})}\BibitemShut {NoStop}%
\bibitem [{\citenamefont {Matsuhira}\ \emph {et~al.}(2002)\citenamefont
  {Matsuhira}, \citenamefont {Hiroi}, \citenamefont {Tayama}, \citenamefont
  {Takagi},\ and\ \citenamefont {Sakakibara}}]{matsuhira2002new}%
  \BibitemOpen
  \bibfield  {author} {\bibinfo {author} {\bibfnamefont {K.}~\bibnamefont
  {Matsuhira}}, \bibinfo {author} {\bibfnamefont {Z.}~\bibnamefont {Hiroi}},
  \bibinfo {author} {\bibfnamefont {T.}~\bibnamefont {Tayama}}, \bibinfo
  {author} {\bibfnamefont {S.}~\bibnamefont {Takagi}}, \ and\ \bibinfo {author}
  {\bibfnamefont {T.}~\bibnamefont {Sakakibara}},\ }\href
  {https://doi.org/10.1088%2F0953-8984%2F14%2F29%2F101} {\bibfield  {journal}
  {\bibinfo  {journal} {Journal of Physics: Condensed Matter}\ }\textbf
  {\bibinfo {volume} {14}},\ \bibinfo {pages} {L559} (\bibinfo {year}
  {2002})}\BibitemShut {NoStop}%
\bibitem [{\citenamefont {Udagawa}\ \emph {et~al.}(2002)\citenamefont
  {Udagawa}, \citenamefont {Ogata},\ and\ \citenamefont
  {Hiroi}}]{doi:10.1143/JPSJ.71.2365}%
  \BibitemOpen
  \bibfield  {author} {\bibinfo {author} {\bibfnamefont {M.}~\bibnamefont
  {Udagawa}}, \bibinfo {author} {\bibfnamefont {M.}~\bibnamefont {Ogata}}, \
  and\ \bibinfo {author} {\bibfnamefont {Z.}~\bibnamefont {Hiroi}},\ }\href
  {\doibase 10.1143/JPSJ.71.2365} {\bibfield  {journal} {\bibinfo  {journal}
  {Journal of the Physical Society of Japan}\ }\textbf {\bibinfo {volume}
  {71}},\ \bibinfo {pages} {2365} (\bibinfo {year} {2002})}\BibitemShut
  {NoStop}%
\bibitem [{\citenamefont {Isakov}\ \emph {et~al.}(2004)\citenamefont {Isakov},
  \citenamefont {Raman}, \citenamefont {Moessner},\ and\ \citenamefont
  {Sondhi}}]{isakov2004magnetization}%
  \BibitemOpen
  \bibfield  {author} {\bibinfo {author} {\bibfnamefont {S.~V.}\ \bibnamefont
  {Isakov}}, \bibinfo {author} {\bibfnamefont {K.~S.}\ \bibnamefont {Raman}},
  \bibinfo {author} {\bibfnamefont {R.}~\bibnamefont {Moessner}}, \ and\
  \bibinfo {author} {\bibfnamefont {S.~L.}\ \bibnamefont {Sondhi}},\ }\href
  {\doibase 10.1103/PhysRevB.70.104418} {\bibfield  {journal} {\bibinfo
  {journal} {Phys. Rev. B}\ }\textbf {\bibinfo {volume} {70}},\ \bibinfo
  {pages} {104418} (\bibinfo {year} {2004})}\BibitemShut {NoStop}%
\bibitem [{\citenamefont {Fisher}(1961)}]{PhysRev.124.1664}%
  \BibitemOpen
  \bibfield  {author} {\bibinfo {author} {\bibfnamefont {M.~E.}\ \bibnamefont
  {Fisher}},\ }\href {\doibase 10.1103/PhysRev.124.1664} {\bibfield  {journal}
  {\bibinfo  {journal} {Phys. Rev.}\ }\textbf {\bibinfo {volume} {124}},\
  \bibinfo {pages} {1664} (\bibinfo {year} {1961})}\BibitemShut {NoStop}%
\bibitem [{\citenamefont {Frob{\"o}se}\ \emph {et~al.}(1996)\citenamefont
  {Frob{\"o}se}, \citenamefont {Bonnemeier},\ and\ \citenamefont
  {J{\"a}ckle}}]{frobose1996orientational}%
  \BibitemOpen
  \bibfield  {author} {\bibinfo {author} {\bibfnamefont {K.}~\bibnamefont
  {Frob{\"o}se}}, \bibinfo {author} {\bibfnamefont {F.}~\bibnamefont
  {Bonnemeier}}, \ and\ \bibinfo {author} {\bibfnamefont {J.}~\bibnamefont
  {J{\"a}ckle}},\ }\href {https://doi.org/10.1088%2F0305-4470%2F29%2F3%2F005}
  {\bibfield  {journal} {\bibinfo  {journal} {Journal of Physics A:
  Mathematical and General}\ }\textbf {\bibinfo {volume} {29}},\ \bibinfo
  {pages} {485} (\bibinfo {year} {1996})}\BibitemShut {NoStop}%
\bibitem [{\citenamefont {Ghosh}\ \emph {et~al.}(2007)\citenamefont {Ghosh},
  \citenamefont {Dhar},\ and\ \citenamefont {Jacobsen}}]{ghosh2007random}%
  \BibitemOpen
  \bibfield  {author} {\bibinfo {author} {\bibfnamefont {A.}~\bibnamefont
  {Ghosh}}, \bibinfo {author} {\bibfnamefont {D.}~\bibnamefont {Dhar}}, \ and\
  \bibinfo {author} {\bibfnamefont {J.~L.}\ \bibnamefont {Jacobsen}},\ }\href
  {https://link.aps.org/doi/10.1103/PhysRevE.75.011115} {\bibfield  {journal}
  {\bibinfo  {journal} {Physical Review E}\ }\textbf {\bibinfo {volume} {75}},\
  \bibinfo {pages} {011115(R)} (\bibinfo {year} {2007})}\BibitemShut {NoStop}%
\bibitem [{\citenamefont {Lee}\ \emph {et~al.}(2017)\citenamefont {Lee},
  \citenamefont {Oh}, \citenamefont {Han},\ and\ \citenamefont
  {Katsura}}]{lee2017resonating}%
  \BibitemOpen
  \bibfield  {author} {\bibinfo {author} {\bibfnamefont {H.}~\bibnamefont
  {Lee}}, \bibinfo {author} {\bibfnamefont {Y.-t.}\ \bibnamefont {Oh}},
  \bibinfo {author} {\bibfnamefont {J.~H.}\ \bibnamefont {Han}}, \ and\
  \bibinfo {author} {\bibfnamefont {H.}~\bibnamefont {Katsura}},\ }\href
  {\doibase 10.1103/PhysRevB.95.060413} {\bibfield  {journal} {\bibinfo
  {journal} {Phys. Rev. B}\ }\textbf {\bibinfo {volume} {95}},\ \bibinfo
  {pages} {060413(R)} (\bibinfo {year} {2017})}\BibitemShut {NoStop}%
\bibitem [{\citenamefont {Dong}\ \emph {et~al.}(2018)\citenamefont {Dong},
  \citenamefont {Chen},\ and\ \citenamefont {Tu}}]{dong20183}%
  \BibitemOpen
  \bibfield  {author} {\bibinfo {author} {\bibfnamefont {X.-Y.}\ \bibnamefont
  {Dong}}, \bibinfo {author} {\bibfnamefont {J.-Y.}\ \bibnamefont {Chen}}, \
  and\ \bibinfo {author} {\bibfnamefont {H.-H.}\ \bibnamefont {Tu}},\ }\href
  {https://link.aps.org/doi/10.1103/PhysRevB.98.205117} {\bibfield  {journal}
  {\bibinfo  {journal} {Physical Review B}\ }\textbf {\bibinfo {volume} {98}},\
  \bibinfo {pages} {205117} (\bibinfo {year} {2018})}\BibitemShut {NoStop}%
\end{thebibliography}%

\end{document}